\documentclass[12pt,a4paper]{article}
\pdfoutput=1
\usepackage[utf8]{inputenc}
\usepackage[table,dvipsnames]{xcolor}
\usepackage{epsf,amsmath,amssymb,graphicx,dcolumn}
\usepackage{caption}
\usepackage{scalefnt,ulem,pstricks}
\usepackage{booktabs,multirow,tabularx}
\usepackage[colorlinks=true,allcolors={blue!70!black}]{hyperref}
\usepackage{cleveref}
\usepackage{rotating}
\usepackage{microtype}
\usepackage{makecell}
\usepackage{cancel}
\usepackage{afterpage}
\usepackage[titletoc,title]{appendix}
\usepackage[numbers,sort&compress]{natbib}
\usepackage{amsmath,amsfonts,amsthm,bm}
\usepackage[tikz]{bclogo}
\usepackage{subcaption}
\usepackage{tikz-feynman}
\usepackage{mathtools}
\providecommand{\href}[2]{#2}

\newcommand\PhiB{\Phi_{\scriptscriptstyle \rm H}}

\newcommand{\CF}{C_{\mathrm{F}}}
\newcommand{\CA}{C_{\mathrm{A}}}

\newcommand\PhiBJ{\Phi_{\scriptscriptstyle \rm HJ}}

\newcommand{\pt}{p_{\text{\scalefont{0.77}T}}}

\newcommand{\muF}{{\mu_{\text{\scalefont{0.77}F}}}}
\newcommand{\muR}{{\mu_{\text{\scalefont{0.77}R}}}}

\newcommand{\muRy}{{\mu_{\text{\scalefont{0.77}R}}^{(0),y}}}
\newcommand{\muRb}{{\mu_{\text{\scalefont{0.77}R}}^{(0),\alpha}}}
\newcommand{\KF}{K_{\text{\scalefont{0.77}F}}}
\newcommand{\KR}{K_{\text{\scalefont{0.77}R}}}
\newcommand{\KRy}{{K^y_{\text{\scalefont{0.77}R}}}}
\newcommand{\KQ}{{K_{\text{\scalefont{0.77}Q}}}}

\newcommand{\noun}[1]{{\scshape #1}}

\newcommand{\POWHEG}{\noun{Powheg}}

\newcommand{\POWHEGBOXRES}{\noun{Powheg-Box-Res}}

\newcommand{\minlo}{{\noun{MiNLO$^{\prime}$}}}
\newcommand{\minnlo}{{\noun{MiNNLO$_{\rm PS}$}}}

\newcommand{\PYTHIA}[1]{\noun{Pythia{#1}}}

\usepackage{xcolor}

\def\to{\rightarrow}

\def\GeV{\mathrm{GeV}}

\newcommand{\eqn}[1]{Eq.\,(\ref{#1})}
\newcommand{\neqn}[1]{Eqs.~(\ref{#1})}
\newcommand{\fig}[1]{Figure~\ref{#1}}

\newcommand{\tab}[1]{Table~\ref{#1}}
\newcommand{\sct}[1]{Section~\ref{#1}}

\def\citere#1{\mbox{Ref.~\cite{#1}}}
\def\citeres#1{\mbox{Refs.~\cite{#1}}}

\usepackage{etoolbox}
\makeatletter
\patchcmd{\@sect}{#8}{\boldmath #8}{}{}
\let\ori@chapter\@chapter
\def\@chapter[#1]#2{\ori@chapter[\boldmath#1]{\boldmath#2}}
\makeatother

\newcommand{\bbH}{\ensuremath{b\bar{b}H}}

\newcommand{\ytsq}{\ensuremath{y_t^2}}

\newcommand{\ybsq}{\ensuremath{y_b^2}}

\newcommand{\dd}{\mathop{}\!\mathrm{d}}

\newcommand{\nolog}[1]{\left.{#1}(\pt)\right|_{\muF=\KF Q_h}}
\newcommand{\atQ}[1]{#1(Q_h)}

\newcommand{\atpt}[1]{{#1}(\pt)}

\allowdisplaybreaks

\begin{document} 
\begin{flushright}
\vspace*{-1.5cm}
MPP-2024-15
\end{flushright}
\vspace{0.cm}

\begin{center}
{\Large \bf NNLO+PS predictions for Higgs production \\[0.2cm] through bottom-quark annihilation with M{\scalefont{0.77}I}NNLO\boldmath{$_{\text{PS}}$}}
\end{center}

\begin{center}
  {\bf Christian Biello$^{(a)}$}, {\bf Aparna Sankar$^{(a,b)}$},  {\bf Marius Wiesemann$^{(a)}$}, and {\bf Giulia Zanderighi$^{(a,b)}$}

$^{(a)}$ Max-Planck-Institut f\"ur Physik, Boltzmannstraße 8, 85748 Garching, Germany\\
  $^{(b)}$ Physik-Department, Technische Universit\"at M\"unchen, James-Franck-Strasse 1, 85748 Garching, Germany \\

\href{mailto:biello@mpp.mpg.de}{\tt biello@mpp.mpg.de},
\href{mailto:aparna@mpp.mpg.de}{\tt aparna@mpp.mpg.de},
\href{mailto:marius.wiesemann@mpp.mpg.de}{\tt marius.wiesemann@mpp.mpg.de},
\href{mailto:zandri@mpp.mpg.de}{\tt zanderi@mpp.mpg.de}

\end{center}

\begin{center} {\bf Abstract} \end{center}\vspace{-1cm}
\begin{quote}
\pretolerance 10000

We consider Higgs production through bottom-quark annihilation at hadron colliders and 
we calculate next-to-next-to-leading-order (NNLO) corrections in QCD perturbation
theory matched to parton showers (NNLO+PS). To this end, we have adapted 
the M{\scalefont{0.77}I}NNLO{\boldmath$_{\text{PS}}$} method to account for the 
extra scale dependence induced by an overall Yukawa coupling that 
is $\overline{\rm MS}$ renormalized. We compare our results against state-of-the-art 
fixed-order predictions at NNLO as well as resummed predictions at 
next-to-next-to-leading-logarithmic (NNLL) accuracy.

\end{quote}

\parskip = 1.2ex

\section{Introduction}
\label{sec:intro}

The Higgs boson provides one of the cornerstones of the Standard Model (SM) of particle physics. With its discovery about a decade ago~\cite{ATLAS:2012yve,CMS:2012qbp} 
the measurement of the properties of the Higgs boson has become a major quest in the rich physics programme at the Large Hadron Collider (LHC).
The exploration of the Higgs sector is of utmost importance not only in the context of the SM, but also in the search for new-physics phenomena.
So far the characterization of the Higgs coupling to \mbox{top ($t$)} and bottom ($b$) quarks, $W$ and $Z$ bosons, and tau leptons is 
fully consistent with the SM picture \cite{ATLAS:2022vkf,CMS:2022dwd}. However, the coupling measurements are becoming continuously more precise with 
the rapid data taking by the LHC experiments, becoming increasingly more sensitive to small deviations from the SM expectations. At the same time, other couplings that are currently restricted due to large statistical uncertainties are anticipated to become more accessible in the future. An example of this is the self-interaction of the Higgs boson.

The accurate simulation of all relevant Higgs-production and decay modes at the LHC is henceforth a crucial requirement for successfully finding deviations from 
the SM predictions. In this context, the associated production of a Higgs boson with bottom quarks (\bbH{}) plays a special role. Although its total rate (of less than $1$\,pb) 
is only about two percent of the dominant Higgs production mode through gluon-fusion, it is still large enough so that its cross section has to be accounted for in precision measurements of the Higgs
boson at the LHC. On the other hand, the direct detection of a \bbH{} signal (by tagging the bottom quarks) is extremely challenging at the LHC, due to large backgrounds 
and because its rate is substantially reduced by requiring one or two bottom quark tags. Even if the measurement of \bbH{} production would be possible, its main purpose 
of extracting a bottom-quark Yukawa coupling is essentially hopeless due to a large contamination from other production mechanisms that do not contain the bottom-Yukawa coupling, see e.g.\ \citere{Pagani:2020rsg}.
Nevertheless, \bbH{} production (and its precise simulation) is particularly relevant in two further respects. Firstly, it is the main Higgs production mechanism
in beyond-the-Standard-Model (BSM) theories with enhanced bottom-Yukawa coupling, for instance for the production of a heavy Higgs boson 
in a Two-Higgs-Doublet-Model (2HDM) like the Minimal-Supersymmetric-SM (MSSM) with a large value of $\tan \beta$.
More importantly, \bbH{} production is the dominant irreducible background in searches for Higgs-boson pair ($HH$) production in the SM in the most sensitive search channels where at least one Higgs boson 
decays to bottom quarks
(for a review, see \citere{DiMicco:2019ngk}). A reliable modeling of the \bbH{} background to $HH$ 
measurements becomes indispensable at the High-Luminosity phase of the LHC (HL-LHC), when the Higgs-pair production cross section in the SM is expected to be measured with a significance of $3.4\sigma$ ($4.9\sigma$)~\cite{ATLAS:2022faz} in the 
combination of all search channels.\footnote{
  The significance reported in brackets assumes negligible systematic uncertainties.}

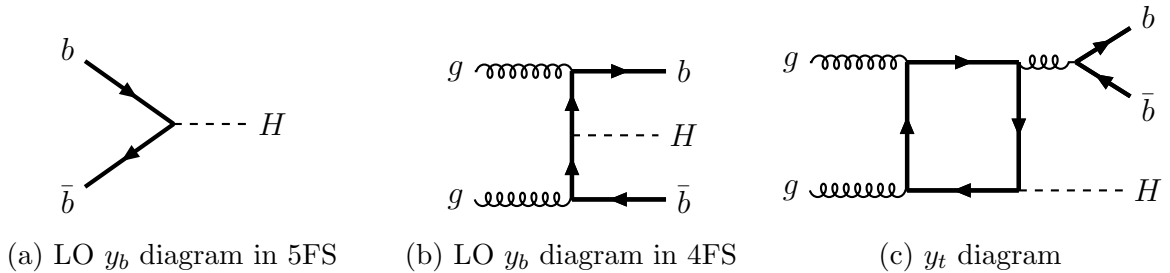
\begin{figure}[t]
  \begin{center}
    \begin{subfigure}[b]{.3\linewidth}
      \centering
\begin{tikzpicture}
\begin{feynman}
	\vertex (a1) at (0,0) {\( b\)};
	\vertex (a2) at (0,-2) {\(\bar b\)};
	\vertex (a3) at (1.4,-1);
	\vertex (a4) at (2.7,-1){\( H\)};
        \diagram* {
          {[edges=fermion]
            (a1)--[ultra thick](a3)--[ultra thick](a2),
          },
          (a3) -- [scalar,thick] (a4),
        };
      \end{feynman}
\end{tikzpicture}
\caption{LO $y_b$ diagram in 5FS}
        \label{subfig:qq}
\end{subfigure}%
\begin{subfigure}[b]{.3\linewidth}
  \centering
\begin{tikzpicture}
  \begin{feynman}
	\vertex (a1) at (0,0) {\( g\)};
	\vertex (a2) at (0,-1.7) {\( g\)};
	\vertex (a3) at (1.53,0);
	\vertex (a4) at (1.53,-1.7);
	\vertex (a42) at (1.53,-0.85);
	\vertex (a43) at (3,-0.85) {\( H\)};
	\vertex (a5) at (3,0){\( b\)};
	\vertex (a6) at (3,-1.7){\(\bar b\)};
        \diagram* {
          {[edges=fermion]
            (a6)--[fermion, ultra thick](a4)--[fermion, ultra thick](a42)--[fermion, ultra thick](a3)--[fermion, ultra thick](a5),
          },
          (a2)--[gluon,thick](a4),
          (a3)--[gluon,thick](a1),
          (a42)--[scalar,thick](a43),
        };
  \end{feynman}
\end{tikzpicture}
\caption{LO $y_b$ diagram in 4FS}
        \label{subfig:gg}
\end{subfigure}%
\begin{subfigure}[b]{.3\linewidth}
  \centering
\begin{tikzpicture}
  \begin{feynman}
	\vertex (a1) at (0,0) {\( g\)};
	\vertex (a2) at (0,-1.7) {\( g\)};
	\vertex (a3) at (1.53,0);
	\vertex (a4) at (1.53,-1.7);
	\vertex (a5) at (3,0);
	\vertex (a6) at (3,-1.7);
	\vertex (a62) at (3.75,0);
	\vertex (a63) at (4.7,0.6){\( b\)};
	\vertex (a64) at (4.7,-0.6){\( \bar b\)};
	\vertex (a7) at (4.7,-1.7) {\( H\)};
        \diagram* {
          {[edges=fermion]
            (a6)--[fermion, ultra thick](a4)--[fermion, ultra thick](a3)--[fermion, ultra thick](a5)--[fermion, ultra thick](a6),
            (a64)--[fermion, ultra thick](a62)--[fermion, ultra thick](a63)
          },
          (a5)--[gluon,thick](a62),
          (a2)--[gluon,thick](a4),
          (a3)--[gluon,thick](a1), 
          (a6)--[scalar,thick](a7),
        };
  \end{feynman}
\end{tikzpicture}\vspace{0.15cm}
\caption{$y_t$ diagram}
        \label{subfig:gg}
\end{subfigure}
\end{center}
\caption{\label{fig:bbh} Sample Feynman diagrams for 
   Higgs production in association with bottom quarks.}
\end{figure}

The dominant contributions to the $b\bar{b}H$ process are those proportional to the bottom Yukawa coupling ($y_b$) where the Higgs couples to a bottom-quark line, see \fig{fig:bbh}\,(a) and (b), as well as those proportional to the top Yukawa coupling ($y_t$) where the Higgs boson couples to a closed top-quark loop, see \fig{fig:bbh}\,(c). In fact, the latter,
which corresponds to the gluon-fusion process with a $b\bar{b}$ pair originating from 
a QCD splitting, has a slightly larger cross-section yield, and its relative size 
further increases when tagging the two bottom quarks. Nevertheless, both production
mechanisms are relevant in the SM and they receive particularly large QCD perturbative corrections, so that higher-order calculations for these processes are crucial.
Other \bbH{} production modes through $VH$ associated production with $V\to b\bar b$ 
and $b$-associated vector-boson fusion have a subleading impact on the cross section 
from a few percent up to a few tenths of percents depending 
on the category, and simulations for these production 
channels exist \cite{Pagani:2020rsg}.

Different schemes can be employed for the calculation of the \bbH{} process, since
the bottom quark can be considered both a massless or a massive quark at 
the typical scale of the $b\bar{b}H$ production at the LHC.
Therefore various predictions have been obtained in both 
a five-flavour scheme (5FS) with massless bottom quarks, 
see \fig{fig:bbh}\,(a) for the respective LO diagram of the process proportional to $y_b$, 
or in a four-flavour scheme (4FS) where the bottom quark is treated as being massive,
see \fig{fig:bbh}\,(b) for a representative LO diagram of the process proportional to $y_b$.
In the 5FS, calculations are technically much simpler and
a significant progress has been made over the past years for the contribution to the cross section proportional to $\ybsq{}$~\cite{Dicus:1998hs,Balazs:1998sb,Harlander:2003ai,Campbell:2002zm,Harlander:2010cz,Ozeren:2010qp,Harlander:2011fx,Buehler:2012cu,Belyaev:2005bs,Harlander:2014hya,Ahmed:2014pka,Gehrmann:2014vha,Duhr:2019kwi,Mondini:2021nck,Wiesemann:2014ioa,Krauss:2016orf,Ajjath:2019ixh,Ajjath:2019neu,Forte:2019hjc,Badger:2021ega},  with the third-order cross sections in QCD being the most remarkable advancement.
In the 4FS, less progress in higher-order calculations has been made 
for the $\ybsq{}$ contribution due to the much more involved structure of the LO process~\cite{Dittmaier:2003ej,Dawson:2003kb,Dawson:2005vi,Liu:2012qu,Dittmaier:2014sva,Zhang:2017mdz,Wiesemann:2014ioa,Jager:2015hka,Krauss:2016orf,Pagani:2020rsg}. Next-to-LO (NLO) corrections in QCD (matched to parton showers) combined with NLO electroweak (EW) corrections are still the state-of-the-art.
On the other hand, a combined study of the $\ybsq{}$ and $\ytsq{}$ contributions to \bbH{} production 
modes has been performed only in the 4FS \cite{Deutschmann:2018avk}, which
included NLO QCD corrections in all relevant coupling structures ($\ybsq{}$, $\ytsq{}$ and $y_b\,y_t$ interference contributions). The matching to parton showers of this 
full NLO QCD calculation has been considered in \citere{Manzoni:2023qaf} and
 studied in the context of backgrounds to $HH$ searches.
 Differences between 4FS and 5FS results have been studied in various works, 
 see e.g.\ \citeres{Maltoni:2012pa,Lim:2016wjo}, and
 consistent combinations of the two schemes have been obtained in  \citeres{Forte:2015hba,Forte:2016sja,Bonvini:2015pxa,Bonvini:2016fgf,Duhr:2020kzd}.

In this paper, we focus on the 5FS calculation of the \bbH{} process proportional to
\ybsq{} and perform the first fully-differential calculation of next-to-NLO (NNLO) QCD corrections
matched to parton showers (NNLO+PS). To this end, we exploit the \minnlo{} method for colour-singlet 
production presented in \citeres{Monni:2019whf,Monni:2020nks} and we adapt the method such that
it can account for an overall scale-dependent 
Yukawa coupling renormalized in the $\overline{\rm MS}$ scheme. We compare our predictions
against reference predictions computed at NNLO in QCD  \cite{Harlander:2012pb,Harlander:2003ai} and against the resummed 
transverse-momentum ($\pt{}$) distribution at NNLL+NNLO accuracy \cite{Harlander:2014hya}.
 
\section{Outline of the calculation}
We consider the production of a Higgs boson through bottom quark annihilation 
\begin{align}
b\bar{b}\to H\,,
\end{align}
in the 5FS, where the bottom quarks are treated as being massless, as given in \fig{fig:bbh}\,(a) for the LO process, which 
is proportional to the bottom-quark Yukawa coupling $y_b$.
From relative order $\alpha_s^2$ on, i.e. NNLO, a loop-induced gluon-fusion contribution proportional to the top-quark Yukawa 
coupling $y_t$ enters, see \fig{fig:bbh}\,(c). We refrain from including this contribution throughout this paper, as it can be considered 
as a completely independent process, which can be obtained via higher-order simulation of Higgs-boson production in gluon fusion.
Moreover, in the 5FS, any $y_b\,y_t$ interference contribution 
between the bottom-quark annihilation and gluon-fusion Higgs production processes
vanish to all orders in perturbation theory, since it is proportional to the bottom-quark mass.

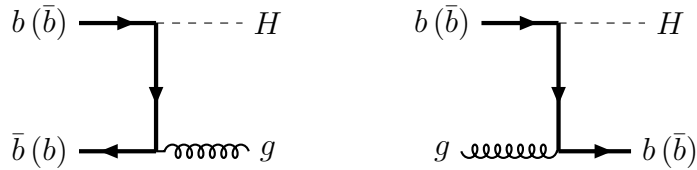
\begin{figure}[t]
  \begin{center}
\begin{subfigure}[b]{.3\linewidth}
  \centering
\begin{tikzpicture}
  \begin{feynman}
	\vertex (a1) at (0,0) {\( b \, (\bar b)\)};
	\vertex (a2) at (0,-1.7) {\( \bar b \, (b)\)};
	\vertex (a3) at (1.53,0);
	\vertex (a4) at (1.53,-1.7);
	\vertex (a5) at (3,0){\( H\)};
	\vertex (a6) at (3,-1.7){\( g \)};
        \diagram* {
          {
            (a6)--[gluon, thick](a4)--[anti fermion, ultra thick](a3)--[scalar](a5),
          },
          (a2)--[anti fermion, ultra thick](a4),
          (a3)--[anti fermion, ultra thick](a1),
        };
  \end{feynman}
\end{tikzpicture}
        \label{subfig:bg}
\end{subfigure}
\begin{subfigure}[b]{.3\linewidth}
  \centering
\begin{tikzpicture}
  \begin{feynman}
	\vertex (a1) at (0,0) {\( b \, (\bar b)\)};
	\vertex (a2) at (0,-1.7) {\( g\)};
	\vertex (a3) at (1.53,0);
	\vertex (a4) at (1.53,-1.7);
	\vertex (a5) at (3,0){\( H\)};
	\vertex (a6) at (3,-1.7){\(b \, (\bar b)\)};
        \diagram* {
          {
            (a6)--[anti fermion, ultra thick](a4)--[anti fermion, ultra thick](a3)--[scalar](a5),
          },
          (a2)--[gluon,thick](a4),
          (a3)--[anti fermion,ultra thick](a1),
        };
  \end{feynman}
\end{tikzpicture}
        \label{subfig:bg}
\end{subfigure}
\end{center}
\caption{\label{fig:bbhj} LO diagrams for $HJ$ production in 5FS.}
\end{figure}

We implement a fully differential computation of Higgs production in bottom-quark annihilation in the 5FS up to NNLO in QCD perturbation theory and consistently match it
to a parton-shower simulation. To this end, we have adapted the \minnlo{} method for colour-singlet production~\cite{Monni:2019whf,Monni:2020nks} to account for an overall scale-dependent 
Yukawa coupling renormalized in the $\overline{\rm MS}$ scheme. And we have implemented an alternative scale setting in all contributions up to NNLO QCD 
that are regular in the transverse momentum of the Higgs boson ($\pt$). Both adaptations of the \minnlo{} method are described in detail in \sct{sec:minnlo}.

The \minnlo{} method has various positive features. In particular, it provides physically sound results without relying on an unphysical slicing scale
to separate events with different jet multiplicities. It is numerically very efficient, since the NNLO corrections are included directly in the event generation without
any post-processing or reweighting of the events. And, not least, it preserves the leading-logarithmic accuracy of the parton shower by keeping the appropriate
ordering of the emissions and scaling intact. The \minnlo{} was not only extended beyond $2\to 1$ production processes in  \citere{Lombardi:2020wju} and applied to several colour-singlet
production processes in \citeres{Lombardi:2020wju,Lombardi:2021rvg,Lombardi:2021wug,Buonocore:2021fnj,Zanoli:2021iyp,Gavardi:2022ixt,Haisch:2022nwz,Lindert:2022qdd,Gauld:2023gtb}, 
it was even reformulated for the case of heavy quark-pair production \cite{Mazzitelli:2020jio,Mazzitelli:2021mmm,Mazzitelli:2023znt}, being the first method applicable to processes with 
colour charges in both initial and final state thus far.

Our \minnlo{} $b\bar{b}\to H$ generator has been implemented within the \POWHEGBOXRES{} framework \cite{Jezo:2015aia}. First, we have implemented a NLO+PS generator for
Higgs plus jet production in bottom-quark annihilation using the \POWHEG{} method \cite{Nason:2004rx,Alioli:2010xd,Frixione:2007vw}, see
\fig{fig:bbhj} for examples of respective LO diagrams. 
For the evaluation of the tree-level amplitudes of the Higgs plus jet ($HJ$) and Higgs plus two-jet $(HJJ)$ processes we 
employ \noun{OpenLoops} \cite{Cascioli:2011va,Buccioni:2017yxi,Buccioni:2019sur}, using its interface \POWHEGBOXRES{}  
developed in \citere{Jezo:2018yaf}. 
For the one-loop virtual corrections to the $HJ$ processes we have used the analytic results from \citere{Harlander:2010cz}, which substantially
improve the numerical performance of the code. We have cross-checked this implementation numerically against \noun{OpenLoops} for several phase space points, finding full agreement at the level
of the machine precision. 
In a second step, we have extended the $HJ$ NLO+PS implementation to NNLO accuracy for $b\bar{b}\to H$ production 
through the newly extended \minnlo{} method described in the next section.

\section{Revising the M{\scalefont{0.77}I}NNLO{\boldmath$_{\text{PS}}$} method}
\label{sec:minnlo}
\subsection{Original method}

In the following, we briefly summarize the \minnlo{} formalism for colour-singlet production, which has been introduced in \citere{Monni:2019whf} and optimized in \citere{Monni:2020nks}. 
The \minnlo{} cross section for $b\bar{b}\to H$ production can be expressed through the standard \POWHEG{} formula 
for $HJ$ production with a modified content of the \POWHEG{} $\bar{B}$ function \cite{Nason:2004rx,Alioli:2010xd,Frixione:2007vw}:
\begin{align}
   \dd \sigma_{\text{\scalefont{0.77}H}}^{\text{\minnlo{}}}= \dd \PhiBJ \bar{B}^{\text{\minnlo{}}} \times \left\{ \Delta_{\text{pwg}}(\Lambda_{\text{pwg}})+ \dd \Phi_{\text{rad}} \Delta_{\text{pwg}}(p_{\text{T,rad}}) \frac{R_{\text{\scalefont{0.77}HJ}}}{B_{\text{\scalefont{0.77}HJ}}} \right\}\,, \label{bpwg}
\end{align}
where $\Phi_{\rm HJ}$ is the $HJ$ phase space.
In the above expression, $\Delta_{\text{pwg}}$ is the \POWHEG{} \,Sudakov form factor with a cutoff $\Lambda_{\text{pwg}}=0.89\text{ GeV}$, while $\dd \Phi_{\text{rad}}$ and $p_{T,\text{rad}}$ are the phase space measure and the transverse momentum of the real radiation with respect to $HJ$ production. In this context, \POWHEG{} takes care of the matching of fixed-order $HJ$ calculation with a parton shower by producing the first additional radiation through the ratio of the tree-level matrix elements for $HJJ$~($R_{\text{\scalefont{0.77}HJ}}$) and $HJ$ productions ($B_{\text{\scalefont{0.77}HJ}}$). All the subsequent radiations with smaller transverse momenta are generated by a Shower Monte Carlo.

The key ingredient of the \minnlo{} method is the modified \POWHEG{} $\bar B$ function which is denoted as $\bar{B}^{\text{\minnlo{}}}$ in \eqn{bpwg}. The derivation of this function stems from the following formula for the NNLO cross section of Higgs production differential in the Born phase space ($\PhiB$) and in the Higgs transverse momentum ($\pt{}$):
\begin{align}
  \frac{\dd \sigma}{\dd \PhiB \dd \pt{}}&=\frac{\dd}{\dd \pt{}}\left\{e^{-\tilde S(\pt{})}\mathcal{L}(\pt{})\right\}+R_f(\pt{}) 
 \nonumber \\&=e^{-\tilde S(\pt{})}\underbrace{\left\{-\mathcal{L}(\pt{})\frac{\dd}{\dd \pt{}}\tilde S(\pt{}) + \frac{\dd}{\dd \pt{}} \mathcal{L}(\pt{})\right\}}_{\eqqcolon\,D(\pt{})}+R_f(\pt{}).\label{masterformula}
\end{align}
In the above equation, the cross section is divided into the singular component and the regular part $R_f(\pt{})$, which is finite in the $p_{T}\rightarrow0$ limit. 
The luminosity $\mathcal{L}(\pt{})$ includes the squared virtual matrix elements for $b\bar{b}\to H$ production and the convolution of the parton densities with 
the collinear coefficient functions ($\tilde C_{ij}$), which can be written as
\begin{align}
  \mathcal{L}(\pt)=\sum_{c\in\{b,\bar b\}} |\mathcal{M}_{c\bar c}^{(0)}|^2 \tilde H_{c\bar c}(\pt) \sum_{ij} \left(\tilde C_{c i} \otimes f_{i}^{[a]}(\pt) \right) \left( \tilde C_{\bar{c} j} \otimes f_j^{[b]} (\pt)\right)\,, \label{lumi}
\end{align}
while the exponent of the Sudakov form factor is defined through
\begin{align}
  \tilde S(\pt{})=2\int_{\pt{}}^M \frac{\dd q}{q} \left\{A(\alpha_s(q))\ln \frac{M^2}{q^2}+\tilde B(\alpha_s(q))\right\}\,, \label{eq:sudakov}
\end{align}
where $M$ is the invariant mass of the colour-singlet final state, i.e.\ $M=m_H$ for Higgs
production in bottom-quark annihilation.
Notice that the hard function ($\tilde H$) and the parton densities are computed at the scale $\pt{}$, which is a feature of the \minnlo{} approach.
In bottom-quark annihilation the Higgs is produced either via $b\bar b$ or $\bar b b$ initial state partons, which is why 
the index $c$ denotes either a bottom quark or antiquark. Since the Born and virtual matrix elements for this process are invariant under charge
conjugation of the initial-state bottom quarks, we will drop the index of $|\mathcal{M}_{c\bar c}^{(0)}|^2$ and $\tilde H_{c\bar c}$ in the following.

The expansion of the relevant resummation coefficients and coefficient functions in the \minnlo{} scheme is given by
\begin{align}
  \tilde H(\pt)&=1+\frac{\alpha_s(\pt)}{2\pi} H^{(1)} + \left( \frac{\alpha_s(\pt)}{2\pi} \right)^2 \tilde H^{(2)} +\mathcal{O}(\alpha_s^3)\,,\label{Hpt} \\
  \tilde C_{ij}(z)&=\delta(1-z) \delta_{ij} + \left(\frac{\alpha_s(\pt)}{2\pi}\right) C^{(1)}_{ij}(z) + \left(\frac{\alpha_s(\pt)}{2\pi} \right)^2 \tilde C_{ij}^{(2)}(z)+\mathcal{O}(\alpha_s^3)\,,\label{eq:Cfunc}\\
  A(\alpha_s)&=\frac{\alpha_s}{2\pi} A^{(1)}+\left(\frac{\alpha_s}{2\pi}\right)^2 A^{(2)}+\left(\frac{\alpha_s}{2\pi}\right)^3 A^{(3)}\,,\\
  \tilde B(\alpha_s)&=\frac{\alpha_s}{2\pi} B^{(1)}+\left(\frac{\alpha_s}{2\pi}\right)^2 \tilde B^{(2)}\,,
\end{align}
with\footnote{The notation $X^{(i)}$ denotes the $i$-th term of $X$ in the perturbative expansion
  in powers of $\frac{\alpha_s}{2\pi}$.}
\begin{align}
  \tilde H^{(2)} &= H^{(2)} -2\zeta_3 A^{(1)} B^{(1)}\,\\
  \tilde C^{(2)}_{ij}(z)&=C^{(2)}_{ij}-2\zeta_3 A^{(1)} \hat P^{(0)}_{ij}(z)\,,\\
  \tilde B^{(2)}&=B^{(2)}+2\pi \beta_0 H^{(1)}+2\zeta_3 \left(A^{(1)}\right)^2\,, \label{b2def}
\end{align}
where $\hat P^{(0)}_{ij}(z)$ is the leading-order regularised splitting function
and $\beta_0=(33-2 n_f)/(12\pi)$, with $n_f$ being the number of light quark flavours.
The latter replacements of the resummation coefficients in the \minnlo{} formalism (marked by the tilde symbol) 
have been derived in detail in \citere{Monni:2019whf}. They originate from the translation from $b$-space to direct space,
while the $H^{(1)}$ term in $\tilde B^{(2)}$ is due to the evaluation of the hard function at the scale $\pt{}$.
The coefficients without tilde are the standard ones for
transverse-momentum resummation production for quark-induced processes, which have been summarized in \citere{Monni:2019whf}.
In particular, for quark-initiated processes $C^{(2)}$ has been obtained in \citere{Catani_2012}, and
the resummation coefficients $A^{(1,2,3)}$ and $B^{(1,2)}$ can be found in \citere{Davies:1984hs}. 

The hard-virtual coefficients $H^{(1)}$ and $H^{(2)}$ are determined from the one-loop and two-loop amplitudes for $b\bar{b}\to H$ production, which have been computed for the 
first time in \citere{Dicus_1999} and \citeres{Bal_zs_1999,Harlander_2003}, respectively. In the \minnlo{} resummation scheme they are expressed as 
\begin{align}
  H^{(1)}&=\CF\left(-2+\frac{7\pi^2}{6}\right)\,,\\
  H^{(2)}&=\CF^2 \left(4-\pi^2 +\frac{67\pi^4}{120}-15\zeta_3\right)+\CF\, n_f \left(\frac{100}{81}-\frac{55\pi^2}{108}+\frac{\zeta_3}{9}\right) \nonumber\\
  &\hspace{0.5cm} +\CA\,\CF \left(-\frac{467}{162}+\frac{701\pi^2}{216}-\frac{2\pi^4}{45}+\frac{151\zeta_3}{18}\right)\,,
\end{align}
where the Casimir factors for $\text{SU}(3)$ are $\CF=\frac{4}{3}$ and $\CA=3$.

We now return to our starting formula in \eqn{masterformula} to derive the \minnlo{} master formula. The regular part $ R_f(\pt)$ can be written as
\begin{align}
  R_f(\pt)=\frac{\dd \sigma_{\text{\scalefont{0.77}HJ}}^{\text{NLO}}(\pt)}{\dd \PhiB \dd \pt}-\frac{\alpha_s(\pt)}{2\pi} \frac{\dd \sigma_s^{(1)}(\pt)}{\dd \PhiB \dd \pt} -\left(\frac{\alpha_s(\pt)}{2\pi}\right)^2 { \frac{\dd \sigma_s^{(2)} (\pt)}{\dd \PhiB \dd \pt}}\,, \label{finitepart}\,
\end{align}
where the first term on the right-hand side of the above equation
is the NLO differential cross section for the production of Higgs boson in association with
one jet ($HJ$) in bottom-quark annihilation, given by
\begin{align}
{\frac{\dd \sigma_{\text{\scalefont{0.77}HJ}}^{\text{NLO}}(\pt)}{\dd \PhiB \dd \pt}}=\frac{\alpha_s(\pt)}{2\pi} {\frac{\dd \sigma_{\text{\scalefont{0.77}HJ}}^{(1)}(\pt)}{\dd \PhiB \dd \pt}}+ \left(\frac{\alpha_s(\pt)}{2\pi}\right)^2
   \frac{\dd \sigma_{\text{\scalefont{0.77}HJ}}^{(2)}(\pt)}{\dd \PhiB \dd \pt}  \label{NLOex}\,.
\end{align}
The notation $X(\mu)$ indicates that the quantity $X$ is evaluated at the scales $\muR=\muF=\mu$. 
Moreover, we can express the coefficients of the expansion of the singular terms as
 \begin{align}
  &{ \frac{\dd \sigma_s^{(1)}(\pt)}{\dd \PhiB \dd \pt}}= \atpt{D^{(1)}},\\
  &{ \frac{\dd \sigma_s^{(2)}(\pt)}{\dd \PhiB \dd \pt}}= \atpt{D^{(2)}}-\tilde S^{(1)}(\pt) \atpt{D^{(1)}}\,.
\end{align}

We can now rewrite \eqn{masterformula} by factoring out the Sudakov exponential and then substitute $R_f(\pt)$ by its expression in \eqn{finitepart}
\begin{align} 
  \frac{\dd \sigma}{\dd \PhiB \dd \pt} &= e^{-\tilde S(\pt)} \left\{ D(\pt) + \frac{R_f(\pt)}{e^{-\tilde S(\pt)}} \right\}\nonumber\\
                                        &= e^{-\tilde S(\pt)} \left\{ D(\pt) + \left(1+\frac{\alpha(\pt)}{2\pi}\tilde S^{(1)}(\pt)\right) R_f(\pt) \right\}+\mathcal{O}(\alpha_s^3)\nonumber \\
                                        &= e^{-\tilde S(\pt)} \left\{ { \frac{\alpha_s(\pt)}{2\pi}\frac{\dd \sigma_{\text{\scalefont{0.77}HJ}}^{(1)}(\pt)}{\dd \PhiB \dd \pt}}\right.  \left(1+\frac{\alpha(\pt)}{2\pi}\tilde S^{(1)}(\pt)\right)   +  \left(\frac{\alpha_s(\pt)}{2\pi}\right)^2{\frac{\dd \sigma_{\text{\scalefont{0.77}HJ}}^{(2)}(\pt)}{\dd \PhiB \dd \pt}}\nonumber \\
                                       & \hspace{2cm}\left.  + \left.\left(D(\pt)-\frac{\alpha_s(\pt{})}{2\pi}\atpt{D^{(1)}}-\left(\frac{\alpha_s(\pt{})}{2\pi}\right)^2 \atpt{D^{(2)}}\right)\right.\right\}+\mathcal{O}(\alpha_s^3)\,. \label{finaldsig}
\end{align}
Factoring out the Sudakov exponential entails two positive features of the \minnlo{} method. Firstly, it improves the numerical stability when integrating over the low-transverse momentum
region. Secondly, it removes the need for a slicing cutoff at small transverse momenta.

Given that we have modified our NNLO accurate starting formula in \eqn{masterformula} only by terms beyond accuracy, our \minnlo{} master formula in \eqn{finaldsig} includes 
NNLO accuracy (upon integration over $\pt{}$) by construction. Notice that up to $\alpha_s^2$ \eqn{finaldsig} corresponds exactly to the \minlo{} formula \cite{Hamilton:2012rf}, which is NLO accurate in
both the $H$ and the $HJ$ phase spaces, while adding the relevant singular terms $\alpha_s^3$ (and beyond), which are 
generated by the total derivative in our starting equation \eqref{masterformula}, adds the relevant contributions required to reach NNLO accuracy in the $H$ phase space upon integration over
$\pt$.

Finally, we apply the same concept to render the \POWHEG{} $\bar B$ function in \eqn{bpwg} NNLO accurate,
by deriving it as
\begin{align}
  \bar{B}^{\text{\minnlo{}}}=e^{-\tilde S(p_{T})} & \left\{ \frac{\alpha_s(\pt)}{2\pi} {\frac{\dd \sigma^{(1)}_{\text{\scalefont{0.77}HJ}}(\pt)}{\dd \Phi_{\text{\scalefont{0.77}HJ}}}}\left(1+\frac{\alpha_s(\pt{})}{2\pi}\tilde S^{(1)}\right)+\left(\frac{\alpha_s(\pt)}{2\pi}\right)^2{\frac{\dd \sigma^{(2)}_{\text{\scalefont{0.77}HJ}}(\pt)}{\dd \Phi_{\text{\scalefont{0.77}HJ}}}}\right. \nonumber\\
                                                  &\quad+\left.\left[D(\pt)-\frac{\alpha_s(\pt{})}{2\pi}\atpt{D^{(1)}}-\left(\frac{\alpha_s(\pt{})}{2\pi}\right)^2 \atpt{D^{(2)}}\right]\times F^{\text{corr}} \right\}\,, \label{bminnlo}
\end{align}
where the function $F^{\text{corr}}$ encodes spreading of the \minnlo{} corrections in the full $\Phi_{\text{\scalefont{0.77}HJ}}$ phase space, given that 
the function $D$ depends only on the kinematical variables of the colour singlet.

At last, we note that the logarithmic terms contained in $\tilde{S}(\pt{})$ and $D(\pt{})$ are switched off at $\pt{}\sim M$ with the following replacement,
\begin{align}
   \log \frac{M}{\pt{}} \rightarrow  \frac{1}{p} \log \left(1+\left( \frac{M}{\pt{}}\right)^p \right),
\end{align}
and taking into account the required jacobian factors. In our calculation, we will use these modified logarithms with $p=6$ in order to match smoothly with the fixed order prediction at high transverse momenta.

\subsection{Accounting for an overall Yukawa coupling}

We now determine the scale dependence of the \minnlo{} formulae and coefficients in the presence of an overall $\overline{\rm MS}$-renormalized Yukawa coupling. 
In Appendix D of \citere{Monni:2019whf}, the scale dependence of the original \minnlo{} formulation was presented for processes with an overall power of the strong coupling at Born level. 
In the case of Higgs production in bottom-quark annihilation in the 5FS, the cross section instead involves two powers of the bottom-Yukawa coupling at Born level. For the sake 
of generality, we present all relevant formulae for a process with the LO coupling structure
\begin{align}
\sigma_{\rm LO}\sim\alpha_s^{n_B}\,y_b^{m_B}\,,
\end{align}
i.e.\ involving both an arbitrary overall power $n_B$ of the strong coupling and an arbitrary overall power $m_B$ of the bottom-Yukawa coupling at Born level.
This is the case for instance for the 4FS process of Higgs production in association with bottom quarks (where $n_B=m_B=2$).
The bottom-quark Yukawa can be expressed through
\begin{align} \label{Yuk}
 y_b = \frac{m_b}{v}\,, 
\end{align}
with $m_b$ being the mass of the bottom quark and $v$ being the vacuum expectation value of the Higgs field.
Below, we present all formulae needed to implement (independent) scale variations in the strong and the Yukawa couplings.

Different renormalization schemes can be employed for the bottom-quark mass in the Yukawa coupling. 
Given that the natural scale of the Yukawa coupling is a hard scale (for instance of the order of the mass of the Higgs boson),
the most appropriate choice is the $\overline{\text{MS}}$ renormalization scheme, which introduces a renormalization scale for 
Yukawa coupling (or more precisely its mass) that can be set to a suitable scale. In the $\overline{\text{MS}}$ scheme,
the scale dependence of Yukawa coupling can be deduced by solving the renormalization group equation (RGE) 
for the bottom-quark mass, which reads 
\begin{align}\label{RGmb}
  &\frac{\dd m_b(\mu)}{m_b(\mu)}=-\gamma(\alpha_s(\mu)) \frac{\dd \mu^2}{ \mu^2}\,.
\end{align}
Here, the anomalous dimension can be expressed as an expansion in $\alpha_s(\mu)$
\begin{align}
  \gamma(\alpha_s(\mu))=\sum_{r=1}^\infty \gamma_r \left(\frac{\alpha_s(\mu)}{2\pi}\right)^r, \,\text{ with } \gamma_1=2 \quad \text{and} \quad \gamma_2=\frac{202}{12}-\frac{20}{36} n_f, 
\end{align}
where we recall that $n_f$ denotes the number of light quark flavours. 
For a generic power $m_B$ of the Yukawa coupling at Born level, we derive the relevant scale-compensating terms up to NNLO
when performing a change of the Yukawa renormalization scale from $M$ to $\muRy$ by solving \eqn{RGmb}
\begin{align}
y_b^{m_B}(M)=y_b^{m_B}(\muRy)&\left\{1 - \frac{\alpha_s(\muRy)}{2\pi} {m_B} \gamma_1 \log \frac{M^2}{(\muRy)^2} - \frac{\alpha_s^2(\muRy)}{\left(2\pi\right)^2}\left[ {m_B} \gamma_2 \log\frac{M^2}{(\muRy)^2}  \right.\right. \nonumber \\
            & \quad \left.\left.+ {m_B} \pi \beta_0  \gamma_1 \log^2 \frac{M^2}{(\muRy)^2}+\frac{{m_B}^2}{2} \gamma_1^2 \log^2 \frac{M^2}{(\muRy)^2} \right] \right\} + {\cal O}(\alpha_s^3)\,.\label{eqyukawaB}
\end{align}
 In the above equation, we introduced an arbitrary scale $\muRy$ for the choice of Yukawa coupling, where the index $(0)$ indicates that this scale is related to the Born cross section. It represents a generic scale that can be varied in order to study the theoretical scale uncertainty related to the Yukawa coupling.

 Accordingly, changing the scale for an arbitrary power $n_B$ of the strong couplings at Born level from $M$ to $\muRb$ takes the following form:
\begin{align}
\frac{\alpha_s^{n_B}(M)}{2\pi}=\frac{\alpha_s^{n_B}(\muRb)}{2\pi}&\left\{1- \frac{\alpha_s(\muRb)}{2\pi} n_B 2 \pi \beta_0 \log\frac{M^2}{(\muRb)^2}+\frac{\alpha_s^2(\muRb)}{\left(2\pi\right)^2}\left[ \frac{1}{2} n_B 4 \pi^2 \beta_0^2  \log^2 \frac{M^2}{(\muRb)^2}  \right.\right. \nonumber \\
	&\quad \left. \left. - n_B 4 \pi^2 \beta_1  \log \frac{M^2}{(\muRb)^2} +\frac{1}{2} n_B^2 4 \pi^2 \beta_0^2  \log^2 \frac{M^2}{(\muRb)^2} \right]\right\}+ {\cal O}(\alpha_s^3)\,,\label{eqalphaB}
\end{align}
with
\begin{align}
   \beta_0=\frac{33-2 n_f}{12\pi},\, \beta_1=\frac{153-19 n_f}{24 \pi^2}.
\end{align}
Here, we introduced an arbitrary scale $\muRb$ for the strong coupling constant in the Born cross section. Also in this case, $\muRb$  can be varied in order to probe 
the uncertainties related to missing higher-order terms.

Within \minnlo{}, the scale-compensating terms originating from the variation of the overall Born couplings are implemented at the level of the hard-virtual 
coefficient function in \eqn{Hpt}. Introducing explicitly the scales $\muRb$ and $\muRy$, the squared virtual matrix element can be written as 
\begin{align}
\begin{split}
  |\mathcal{M}|^2&=|\mathcal{M}^{(0)}(M,M)|^2\,\left(1+\frac{\alpha_s(\pt{})}{2\pi}H^{(1)}+\left(\frac{\alpha_s(\pt)}{2\pi}\right)^2 \tilde H^{(2)}\right) \\
                 &=|\mathcal{M}^{(0)}(\muRb,\muRy)|^2\,\left(1+\frac{\alpha_s(\muR)}{2\pi}H^{(1)}(\KR,\tfrac{\muRb}{M},\tfrac{\muRy}{M}) \right. \\
  &\hspace{4.85cm}+\left.\left(\frac{\alpha_s(\muR)}{2\pi}\right)^2 \tilde H^{(2)}(\KR,\tfrac{\muRb}{M},\tfrac{\muRy}{M})\right)+{\cal O}(\alpha_s^3)\,. \label{Hwithscales}
    \end{split}
\end{align}
Here, $\mathcal{M}^{(0)}(\muRb, \muRy)$ is the tree-level amplitude with the strong and Yukawa couplings evaluated at $\muRb$ and $\muRy$, respectively. 
In addition, we have introduced a generic symbol $\muR$ for the renormalization scale of the extra powers of the strong coupling in the expansion of $H$,
which is set to $\muR=K_R\,\pt{}$ following the \minnlo{} prescription. Using the identity 
\begin{align}
  |\mathcal{M}^{(0)}(M,M)|^2=|\mathcal{M}^{(0)}(\muRb,\muRy)|^2 \frac{\alpha_s^{n_B}(M)y^{{m_B}}_b(M)}{\alpha_s^{(n_B)}(\muRb) y^{m_B}(\muRy)},
\end{align}
while inserting the relations in \neqn{eqyukawaB} and \eqref{eqalphaB}, we can absorb the logarithmic scale-compensating terms into the hard-virtual coefficient function
\begin{align}
  H^{(1)}(\KR,\tfrac{\muRb}{M},\tfrac{\muRy}{M})=&\,H^{(1)}+ n_B 2\pi\beta_0 \log\frac{(\muRb)^2}{M^2}+m_B \gamma_1 \log \frac{(\muRy)^2}{M^2}, \label{H1yuk}\\
	\tilde H^{(2)}(\KR,\tfrac{\muRb}{M},\tfrac{\muRy}{M})=&\,\tilde H^{(2)}+ \left(2\pi\beta_0 \ln K_R^2+n_B 2\pi\beta_0 \log\frac{(\muRb)^2}{M^2}+m_B \gamma_1 \log \frac{(\muRy)^2}{M^2}\right)H^{(1)}\nonumber\\
	&+n_B 4 \pi^2 \beta_1  \log \frac{(\muRb)^2}{M^2} +\frac{1}{2} n_B(n_B-1) 4 \pi^2 \beta_0^2  \log^2 \frac{(\muRb)^2}{M^2} \nonumber\\
	&+n_B 4\pi^2 \beta_0^2 \log\frac{(\muRb)^2}{M^2} \log \KR^2 +n_B 2\pi\beta_0 m_B \gamma_1\log \frac{(\muRy)^2}{M^2} \log\frac{(\muRb)^2}{M^2}     \nonumber\\
        &+ m_B \gamma_2 \log\frac{(\muRy)^2}{M^2} {- m_B \pi \beta_0  \gamma_1 \log^2 \frac{(\muRy)^2}{M^2}} + \frac{1}{2} m_B^2 \gamma_1^2 \log^2 \frac{(\muRy)^2}{M^2} \nonumber  \\
  &{+ m_B 2\pi\beta_0 \gamma_1 \log \frac{(\muRy)^2}{M^2} \log \KR^2}. \label{H2yuk}
\end{align}
For Higgs production in bottom-quark annihilation 
the coefficients $\beta_{0,1}$ and $\gamma_{1,2}$ are calculated considering five massless quarks ($n_f=5$), and there is no strong coupling ($n_B=0$), but two powers of 
Yukawa couplings (${m_B}=2$) at the Born level. Moreover, the invariant mass $M$ corresponds to $m_H$ for this process. As a result of the modification of $H^{(1)}$, also the 
$B^{(2)}$ coefficient in the Sudakov receives a $\muRb$ and $\muRy$ dependence through its dependence on $H^{(1)}$ within \minnlo{}, see \eqn{b2def}.
For completeness we provide here also the standard $\muR$ dependence of the Sudakov coefficients, whose complete scale dependence is implemented through
\begin{align}
  & A^{(2)}(\KR)=A^{(2)}+(2\pi \beta_0) A^{(1)} \log \KR^2,\\
  &\tilde B^{(2)}(\KR,\tfrac{\muRb}{M},\tfrac{\muRy}{M})= \,\,\tilde B^{(2)}+(2\pi \beta_0)B^{(1)} \log \KR^2+n_B (2\pi \beta_0)^2 \log \frac{(\muRb)^2}{M^2} \nonumber\\
  &\hspace{4.4cm}+m_B 2 \pi \beta_0 \gamma_1 \log \frac{(\muRy)^2}{M^2}.
\end{align}
To assess the theoretical uncertainty of our \minnlo{} predictions, we can now vary $\muR$, $\muRb$ and $\muRy$ around their central scales, either simultaneously by a common 
factor or independently. Our default choice and the impact on Higgs production in bottom-quark annihilation will be discussed in detail in \sct{sec:results}.

In \citere{Mazzitelli:2021mmm}, a scale $Q=\KQ M$ in the modified logarithm was introduced, dubbed resummation scale, which determines the region at large transverse momentum 
where resummation effects are smoothly turned off within \minnlo{}.  Since there is an interplay with scale of the Yukawa coupling $\muRy$, we report the full scale dependence 
for $\KQ$, $\muR$, $\muRb$ and $\muRy$ of the hard-virtual coefficient function below.
The resummation-scale dependence is derived by splitting the integral of the Sudakov form factor in \eqn{eq:sudakov} into one contribution from $\pt$ to $Q$ and a second 
one from $Q$ to $M$, where the second integral is then expanded in powers of $\alpha_s(\KR/\KQ\,\pt)$ up to second order.
While the logarithmic contributions are absorbed into a redefinition of the $\tilde B^{(2)}$ coefficient, see Eq.\,(4.8) of \citere{Mazzitelli:2021mmm}, the 
non-logarithmic terms are expanded outside the exponential factor and are absorbed into the hard-virtual coefficient function $H$. 
For consistency the scale of the strong coupling in the expansion of $H$ is changed as follows:
\begin{align}
  \muR=\KR \pt \rightarrow \muR=\frac{\KR}{\KQ}\pt\,,
\end{align}
while the complete scale dependence of the expansion coefficients of $H$ reads
\begin{align}
  &H^{(1)}(\KR,\tfrac{\muRb}{M},\tfrac{\muRy}{M},\KQ)= H^{(1)}(\KR,\tfrac{\muRb}{M},\tfrac{\muRy}{M})+\left(-\frac{A^{(1)}}{2}\log\KQ^2+B^{(1)}\right) \log \KQ^2,\\
  &\tilde H^{(2)}(\KR,\tfrac{\muRb}{M},\tfrac{\muRy}{M},\KQ)= \tilde H^{(2)}(\KR,\tfrac{\muRb}{M},\tfrac{\muRy}{M})+\frac{(A^{(1)})^2}{8}\log^4 \KQ^2 \nonumber\\
                                                    &\hspace{1cm}-\left(\frac{A^{(1)}B^{(1)}}{2}+\pi\beta_0\frac{A^{(1)}}{3}\right)\log^3 \KQ^2+\left(-\frac{A^{(2)}(\KR)}{2}+\frac{(B^{(1)})^2}{2}+\pi\beta_0 B^{(1)}\right. \nonumber\\
                                                    &\hspace{1cm} \left.-n_B \pi \beta_0 A^{(1)} \log \frac{(\muRb)^2}{M^2} -\frac{1}{2}m_B \gamma_1 A^{(1)} \log \frac{(\muRy)^2}{M^2} \right)\log^2 \KQ^2 +\left(\tilde B^{(2)}(\KR,\tfrac{\muRb}{M},\tfrac{\muRy}{M})\right.\nonumber \\
                                                    &\hspace{1cm} \left.+2 n_B \pi \beta_0 B^{(1)} \log \frac{(\muRb)^2}{M^2}+ m_B \gamma_1 B^{(1)} \log \frac{(\muRy)^2}{M^2}\right) \log \KQ^2 + \left( B^{(1)} \log \KQ^2 \right.\nonumber \\
                                                    &\hspace{1cm}  \left.-\frac{A^{(1)}}{2}\log^2 \KQ^2 -2\pi\beta_0 \log \KQ^2\right) H^{(1)}(\KR,\tfrac{\muRb}{M},\tfrac{\muRy}{M}).
\end{align}
We note that we refrained from reporting the factorization scale ($\muF=\KF\,\pt$) dependence of the \minnlo{} method here, which is absorbed into the collinear 
coefficient functions, see \eqn{eq:Cfunc}, as there is no direct interplay with $\muRy$. Instead, we refer to \citere{Mazzitelli:2021mmm} for the relevant formulae.

\subsection{Alternative scale setting for the non-singular contribution ({\tt FOatQ})}
\label{sec:FOatQ}
In the following, we explore a different scale choice for the non-singular part $R_f$ in \eqn{finitepart}, which can improve the comparison with the fixed-order computations by reducing
differences in the treatment of terms beyond accuracy. In particular, we rederive the \minnlo{} formulae for the case where $R_f$ is evaluated at a hard scale $Q_h$ instead of the transverse momentum $\pt{}$ as used in the original formulation of the \minnlo{} method.\footnote{Such alternative scale setting was also employed for diphoton production in \citere{Gavardi:2022ixt}.} 
The singular part encoded through the $D$ function remains to be evaluated at the scale $\pt{}$. 
More precisely, we will set the scale of the PDFs to $\muF=\KF Q_h$ and the scale of the extra powers of the strong coupling (those beyond the ones appearing at LO) to $\muR=\KR Q_h$,
in order to facilitate the usual scale variations from the central scale $Q_h$ through $\KF$ and  $\KR$.

We start by writing the regular part as
\begin{align}
  \atQ{R_f}=\frac{\atQ{\dd \sigma_{\text{\scalefont{0.77}HJ}}^{\text{NLO}}}}{\dd \PhiB \dd \pt}-\frac{\alpha_s(\KR Q_h)}{2\pi} \frac{\atQ{\dd \sigma_s^{(1)}}}{\dd \PhiB \dd \pt} -\left(\frac{\alpha_s(\KR Q_h)}{2\pi}\right)^2 \frac{\atQ{ \dd \sigma_s^{(2)}}}{\dd \PhiB \dd \pt}\,.
\end{align}
Here, the argument of $R_f$ being ``$(Q_h)$" simply denotes that $\muR/\KR=\muF/\KF=Q_h$ as opposed to  ``$(\pt)$", which implies $\muR/\KR=\muF/\KF=\pt$.
As a result and to keep the singular terms evaluated at a scale $\pt$ the $\bar{B}$ function has to be modified as
\begin{align}
  \bar{B}^{\text{\minnlo{}}}=&e^{-\tilde S(p_{T})} \left\{ \frac{\alpha_s(\KR Q_h)}{2\pi}\frac{\atQ{\dd \sigma^{(1)}_{\text{\scalefont{0.77}HJ}}}}{\dd \Phi_{\text{\scalefont{0.77}HJ}}}\left(1+\frac{\alpha_s(\KR Q_h)}{2\pi}\tilde S^{(1)}\right)+\left(\frac{\alpha_s(\KR Q_h)}{2\pi}\right)^2\frac{\atQ{\dd \sigma^{(2)}_{\text{\scalefont{0.77}HJ}}}}{\dd \Phi_{\text{\scalefont{0.77}HJ}}}\right. \nonumber\\
  &\left.+\left(D(\pt)-\frac{\alpha_s(\KR Q_h)}{2\pi}\atQ{D^{(1)}}-\left(\frac{\alpha_s(\KR Q_h)}{2\pi}\right)^2 \atQ{D^{(2)}}\right)\times F^{\text{corr}} \right\}.\label{bminnlofoatq}
\end{align}
We note that, as opposed to the implementation used in \citere{Gavardi:2022ixt}, we perform a factorization of the full Sudakov form factor, also in front of the regular part.
By contrast, \citere{Gavardi:2022ixt} employed a less accurate form of the Sudakov form factor, denoted as $\bar{S}$ in Eq.\,(2.30) of that paper, to be factored out in front of the regular part.
The two approaches are equivalent up to the desired accuracy of the \minnlo{} method. 
However, to us it seems more natural and practically simpler to use the same exponential factor for both the singular
and the regular part.

This implementation of the scale choice can be turned on through setting the flag \texttt{FOatQ\,1} in the code. The two new ingredients 
that are required to implement \eqn{bminnlofoatq} are $D^{(1)}(Q_h)$ and  $D^{(2)}(Q_h)$, whose formulae we will derive and provide in the following.
They are defined by writing the expansion of $D$ with a scale setting of $\muR/\KR=\muF/\KF=Q_h$\footnote{We note that we only require the first order of the DGLAP/RGE expansions here,
as any further terms would enter beyond the desired accuracy. Also for that reason, we have changed the argument of the strong coupling in the expansion of the DGLAP evolution to be
$\KR Q_h$ rather than $\KF Q_h$, since again the difference is beyond the desired accuracy of \minnlo{}.}
\begin{align}
  \atQ{D} = \frac{\alpha_s(\KR Q_h)}{2\pi} \atQ{D^{(1)}} + \left( \frac{\alpha_s(\KR Q_h)}{2\pi} \right)^2 \atQ{D^{(2)}} + \mathcal{O}(\alpha_s^3)\,,
\end{align}
whereas the corresponding expression for $\muR/\KR=\muF/\KF=\pt$ reads
\begin{align}
  D(\pt) = \frac{\alpha_s(\KR \pt)}{2\pi} D^{(1)}(\pt{}) + \left( \frac{\alpha_s(\KR \pt)}{2\pi} \right)^2 D^{(2)}(\pt) + \mathcal{O}(\alpha_s^3)\,.
\end{align}
To derive $\atQ{D}$ we simply start from $D(\pt)$ and consistently change the scale setting in the parton distribution functions (PDFs) and the strong coupling. To this end, we employ
the first-order expansion of the DGLAP evolution of the PDFs
\begin{align}
  f_c^{[a]}(\pt)  = f_c^{[a]}(\KF Q_h)+\frac{\alpha_s(\KR Q_h)}{2\pi}\sum_i \left(\hat P_{ci}^{(0)}\otimes f_i^{[a]}\right)\log\frac{\pt^2}{\KF^2 Q_h^2}+\mathcal{O}(\alpha_s^2)\,,\label{eq:DGLAP}
\end{align}
and of the RGE evolution of the strong coupling
\begin{align}
  \alpha_s(\pt)=\alpha_s(\KR Q_h)\left[1-\frac{\alpha_s(\KR Q_h)}{2\pi}\, 2\,\pi\,\beta_0 \log \frac{\pt^2}{\KR^2 Q_h^2}\right]+\mathcal{O}(\alpha_s^3)\,.\label{eq:RGEalpha}
\end{align}

Note that the function $D$ is defined in a scale invariant way, so that $\atQ{D}$ and $\atpt{D}$ is simply the same function, but with a different scale choice, and they differ
only by terms beyond accuracy. 
When imposing the expansion of the DGLAP and the RGE evolution in \eqn{eq:DGLAP} and \eqref{eq:RGEalpha}, respectively, on the first-order coefficient $D^{(1)}$ to change 
its scale from $\pt$ to $Q_h$, the logarithmic scale dependence is absorbed into the second-order coefficient $D^{(2)}$. Thus, the expansion coefficients of $D(Q_h)$ can be expressed as
\begin{align}
  \atQ{D^{(1)}}=&\hspace{0.2cm}\nolog{D^{(1)}}\label{D1}\\
  \atQ{D^{(2)}}=&\hspace{0.2cm}\nolog{D^{(2)}}+ \left[\hat P^{(0)} \otimes D^{(1)}\right]_{\text{symb}}  \log \frac{\pt^2}{Q_h^2}-2\pi\beta_0 \atQ{D^{(1)}} \log\frac{\pt^2}{Q_h^2},  \label{D2withlog}
\end{align}
where the notation $\nolog{D^{(i)}}$ simply denotes to take the exact functional form of $D^{(i)}(\pt)$, but setting the scale of the PDFs to $\KF Q_h$.
Note that $\nolog{D}$ is not scale independent up to higher orders, while the function $D(Q_h)$ is scale invariant due to the inclusion of the logarithmic scale-compensating terms.
Moreover, we have introduced a symbolic shortcut in \eqn{D2withlog} for the effect of the convolution in the DGLAP evolution of the PDFs in $D^{(1)}$ whose explicit form reads
\begin{align}
             \left[\hat P^{(0)} \otimes D^{(1)}\right]_{\text{symb}} \coloneqq&\hspace{0.2cm} -\left[\frac{\atQ{\dd \tilde S^{(1)}}}{\dd \pt}\right] \sum_{c\in{\{b,\bar b\}}} \left|\mathcal{M}^{(0)}\right|^2 \sum_i \left[f^{[a]}_c \left(\hat P^{(0)}_{\bar c i} \otimes f^{[b]}_{i}\right)+\left(\hat P^{(0)}_{ci}\otimes f^{[a]}_{i}\right) f^{[b]}_{\bar c} \right] + \nonumber\\
             &\hspace{0.2cm}\sum_{c\in{\{b,\bar b\}}} \left|\mathcal{M}^{(0)}\right|^2 \sum_{ij} \left[ f^{[a]}_c \left( \hat P^{(0)}_{\bar c i}\otimes \hat P^{(0)}_{ij} \otimes f^{[b]}_{j}\right) + \left(\hat P^{(0)}_{ci} \otimes \hat P^{(0)}_{ij} \otimes f^{[a]}_j\right)f^{[b]}_{\bar c}  \right. \nonumber\\
               &\hspace{0.2cm}\left.+2 \left(\hat P^{(0)}_{ci} \otimes f^{[a]}_{i}\right)\left(\hat P^{(0)}_{\bar c j} \otimes f_j^{[b]}\right) \right]. \label{symbconv}
\end{align}
Lastly, we only need to specify the relevant formulae to determine $D^{(1)}(\pt)$ and $D^{(2)}(\pt)$, for which we refer the reader 
to Eq.\,(27) and (28) of \citere{Monni:2020nks} and to \citere{Mazzitelli:2021mmm} for the relevant $\KR$ and $\KF$ dependence. 
We stress that, apart from the formulae specified above, all scale dependent terms in  $\KR$ and $\KF$ remain exactly as in the original 
formulation of \minnlo{}.

Our notation here, in particular the simplicity of \neqn{D1} and \eqref{D2withlog}, reflects exactly how the modifications for {\tt FOatQ} have been implemented 
in the code. 
The implementation of the {\tt FOatQ} option has been performed  within the \POWHEGBOXRES{} code, which will be included in a future release, 
so that this feature can be used for all publicly available \minnlo{} codes.
We have performed additional checks of our \texttt{FOatQ} implementation by comparing both prescriptions for Drell-Yan production and for Higgs-boson production in gluon fusion.

\section{Results}
\label{sec:results}

We present numerical predictions for Higgs production in bottom-quark annihilation ($b\bar{b}\to H$) for the LHC at 13\,TeV centre-of-mass energy.
We choose the following input parameters: For the (stable) Higgs boson the mass is set to $m_H = 125$\,GeV and its width to $\Gamma_H = 0$\,GeV.
The 5FS is used with massless bottom quarks, but with a non-vanishing Yukawa coupling $y_b$, renormalized in the $\overline{\rm \text{MS}}$ scheme.
The Yukawa coupling is evaluated from an input value $m_b(m_b) =4.18\, \GeV$ and evolved to its respective central scale $\muRy=m_H$ using
four-loop running, while scale variations are obtained from that central value with a three-loop evolution, consistent with the order of our calculation. 
This procedure follows the recommendation by the LHC Higgs cross section working group~\cite{deFlorian:2016spz}.
We use the NNLO set of NNPDF40~\cite{NNPDF:2021njg} parton densities with 5 active flavours
 with \mbox{$\alpha_s(m_Z)$ = 0.118} via the LHAPDF interface~\cite{Buckley:2014ana} as our default setting unless specified otherwise.
The central factorization and renormalization scales are set following to the \minnlo{} method~\cite{Monni:2019whf}, and we also present results with the 
{\tt FOatQ 1} scale setting discussed in \sct{sec:FOatQ}.
The associated scale uncertainties are determined 
through the customary 7-point envelope obtained by varying independently the corresponding factors $\KR$ and $\KF$ by a factor 2, where the scale
of $y_b$ is varied simultaneously with $\KR$, i.e.\ $\muRy=\KRy m_H$ where $\KRy=\KR$.
Moreover, we choose $\KQ=0.25$, while we have checked that setting $\KQ=0.5$ leads to rather small differences in the results.
These scale settings are in line with ones used for the analytic resummation for Higgs production in bottom-quark annihilation 
presented in \citere{Harlander:2014hya}, which we will compare to at the end of this section.

For the predictions matched to a parton shower we employ {\sc Pythia8} with the A14 tune (\texttt{py8tune 21} 
in the input card). Since we are interested in inclusive Higgs production and comparisons to other theory calculations, 
the Higgs boson is kept stable, and the effects from hadronization, 
multi-parton interactions (MPI) and QED radiation are kept off.

{\renewcommand{\arraystretch}{1.5}
\begin{table*}[htp!]
  \vspace*{0.3ex}
  \begin{center}
\begin{small}
    \begin{tabular}{||c|c|c|c|c|c||}
\hline\hline
Process & NLO ({\sc SusHi}) & NNLO ({\sc SusHi}) & \minlo{} & \minnlo{} & \begin{tabular}{@{}c@{}} \minnlo{} \vspace{-2mm}\\  ({\tt FOatQ 1})\end{tabular} \\
\hline\hline
$b\bar b \rightarrow H $ & $0.646(0)_{-10.9\%}^{+10.4\%}$\,pb & $0.518(2)_{-7.5\%}^{+7.2\%}$\,pb & $0.571(1)_{-22.7\%}^{+17.4\%}$\,pb & $0.509(8)_{-5.3\%}^{+2.9\%}$\,pb & $0.508(4)_{-4.3\%}^{+3.6\%}$\,pb \\
\hline\hline
    \end{tabular}
\end{small}
  \end{center}
  \caption{Total cross sections of Higgs-boson production in bottom-quark annihilation. The number in brackets denotes the numerical uncertainty on the last digit, while the uncertainty quoted is the theory uncertainty estimated via scale variations as described in the text.}
\label{tab:H_xs}
\end{table*}
}

We start the discussion of our phenomenological predictions by comparing predictions 
for the total inclusive cross section from \minlo{} and \minnlo{} with  
fixed-order results at NLO and NNLO in \tab{tab:H_xs}. The fixed-order rates for $b\bar{b}\to H$ production
are obtained with the program {\sc SusHi} \cite{Harlander:2012pb,Harlander:2003ai} and they have been obtained 
with the input parameters specified above, while setting renormalization and factorization scales to $\muR/\KR=\muF/\KF=m_H$. For \minnlo{} we quote the 
results for both the standard and the {\tt FOatQ 1} scale setting, and find their total $bb\to H$ cross sections to be very close. We further observe that
the effect of NNLO QCD corrections, both when comparing the fixed-order numbers at NNLO QCD with NLO QCD from {\sc SusHi} and 
when comparing \minnlo{} to \minlo{}, is to reduce the cross section by more than 10\%. Moreover, we observe a significant reduction of scale uncertainties
once NNLO QCD corrections are included. Finally, our \minnlo{} predictions are in agreement with the NNLO QCD cross section within the quoted 
uncertainties from missing higher-order corrections. We stress that due to the different scale setting and a different treatment of terms beyond 
accuracy \minnlo{} predictions are not expected to coincide with NNLO QCD results, but rather to agree
with them within the given uncertainties.

Although all the results in this paper are obtained with the correlated scale variation ($\KRy=\KR$) for the Yukawa and strong couplings, we implemented in the code the possibility to change the two renormalisation factors in an independent way in order to study the Yukawa effects. If we vary $\KRy$ and $\KF$ keeping $\KR=1$, we observe a more symmetric scale variation for the total cross-section, precisely $0.509(8)_{-5.1\%}^{+4.7\%}$\,pb with the standard settings without \texttt{FOatQ 1}.

\begin{figure}[t!]
\begin{center}
\includegraphics[width=.52\textwidth, page=1]{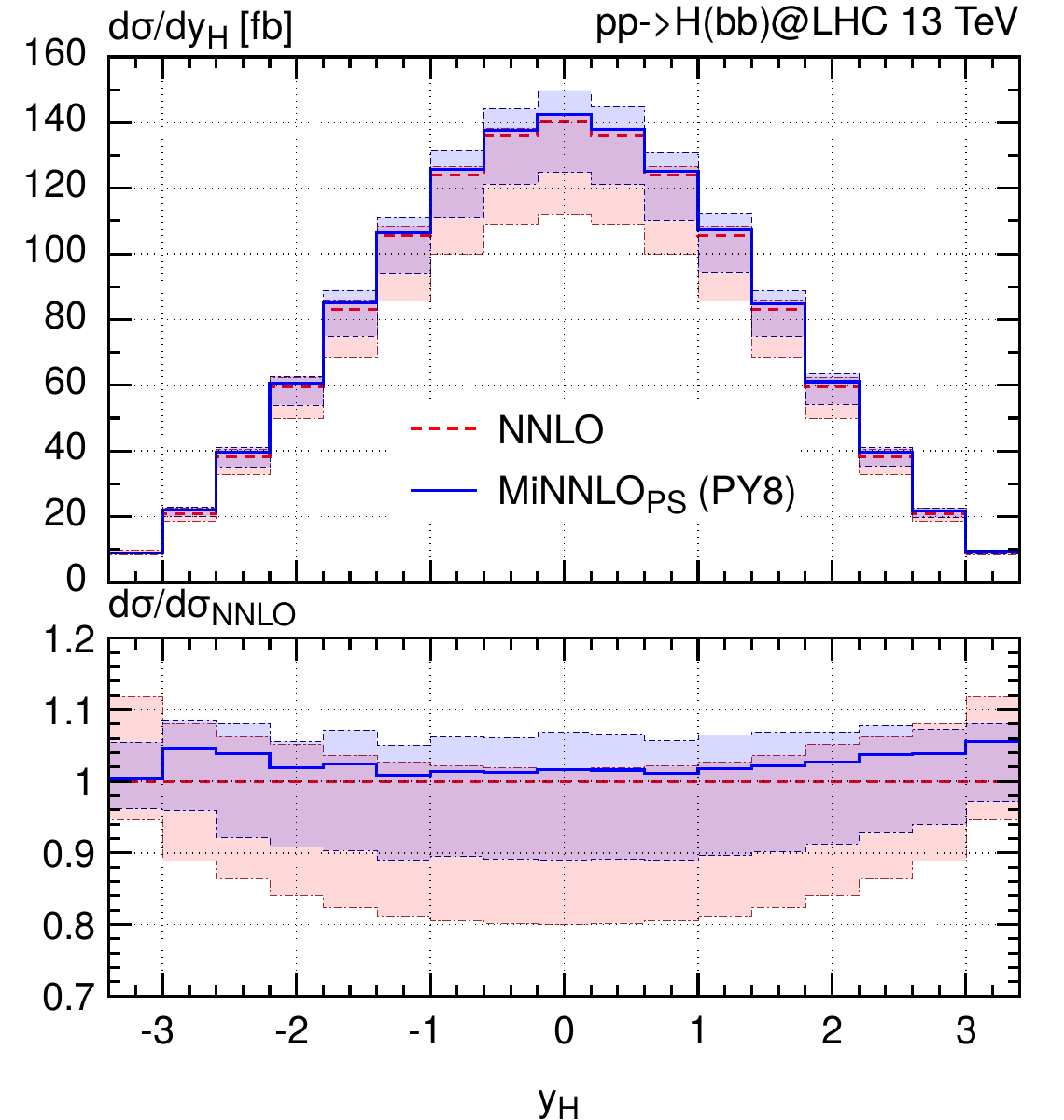}
\vspace*{2ex}
\caption{Comparison of \minnlo{} predictions (blue, solid) with the NNLO results of \citere{Mondini:2021nck}
(red, dashed) for the rapidity distribution of the Higgs boson.  \label{fig:yHNNLO}}
\end{center}
\end{figure}

\begin{figure}[t!]
\begin{center}
\begin{tabular}{cc}
\includegraphics[width=.45\textwidth]{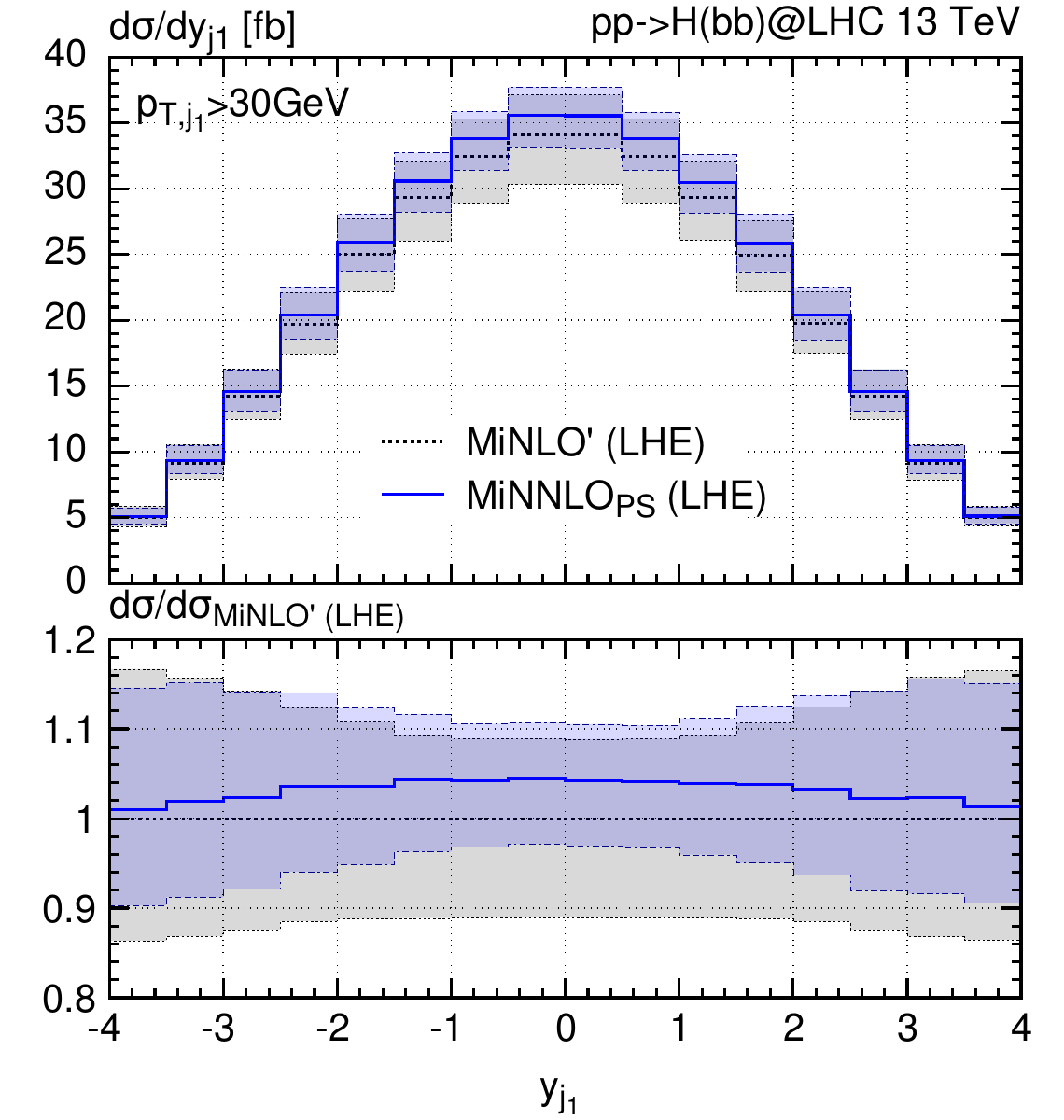}
&
\hspace{-0.8cm}
\includegraphics[width=.45\textwidth]{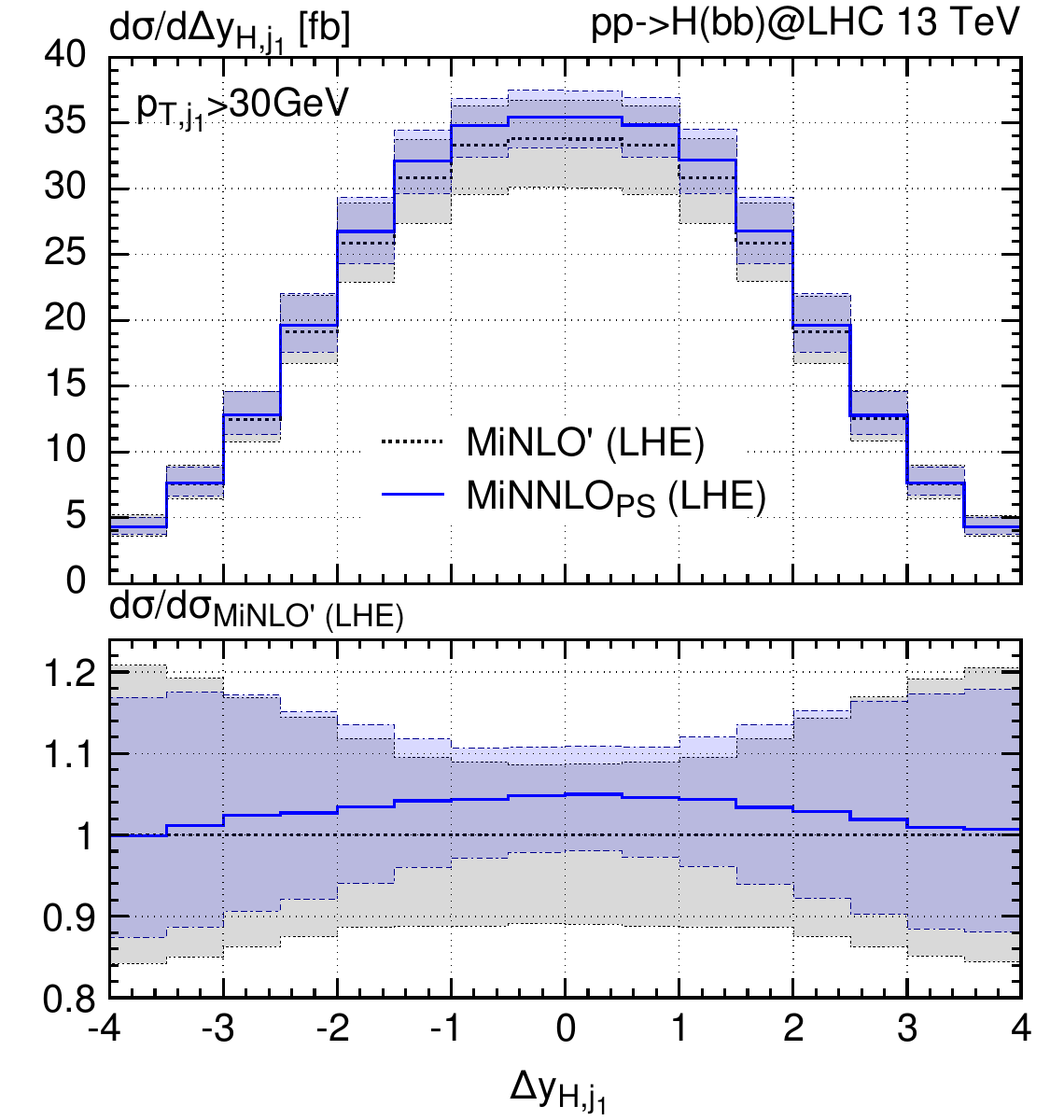}\\
\includegraphics[width=.45\textwidth]{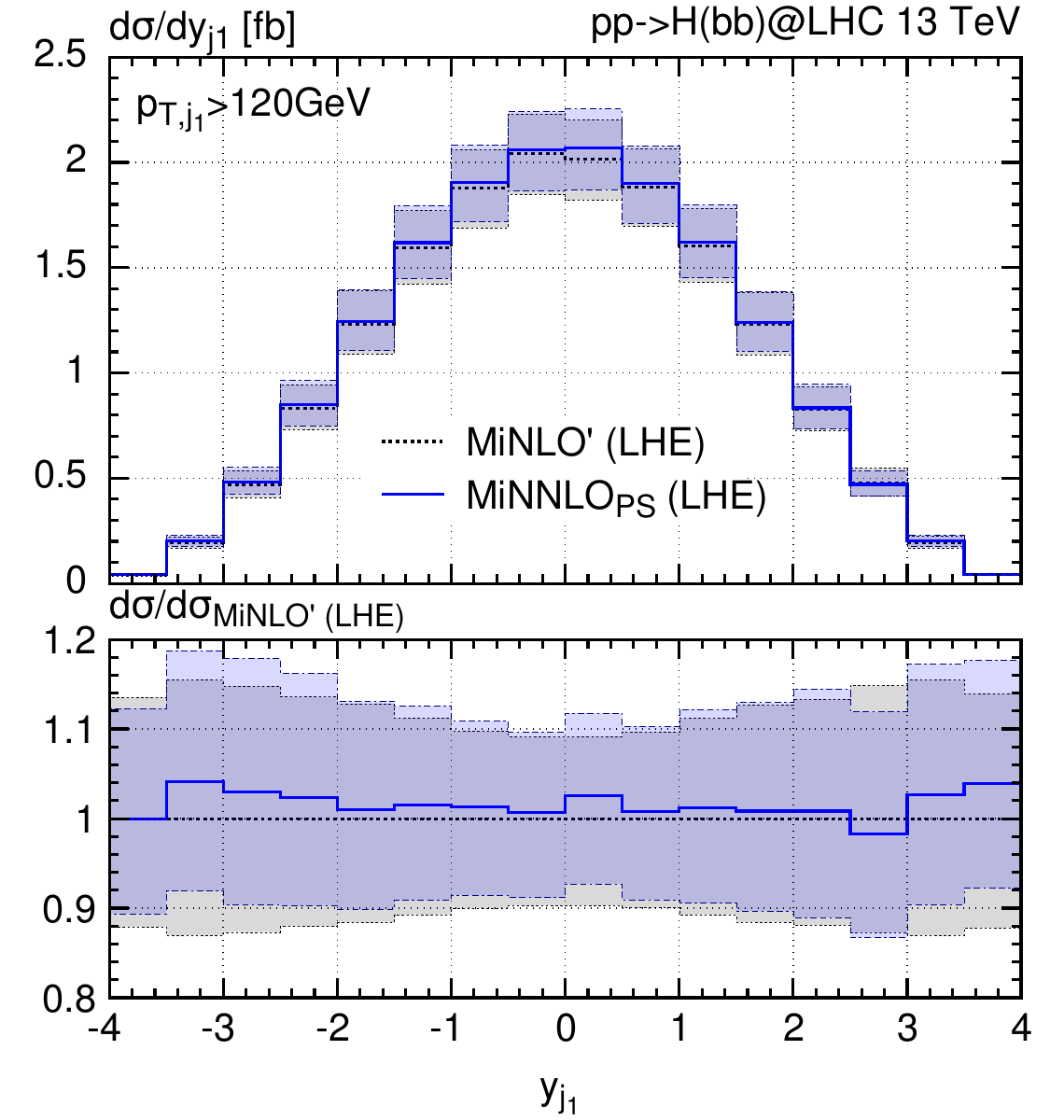}
&
\hspace{-0.8cm}
\includegraphics[width=.45\textwidth]{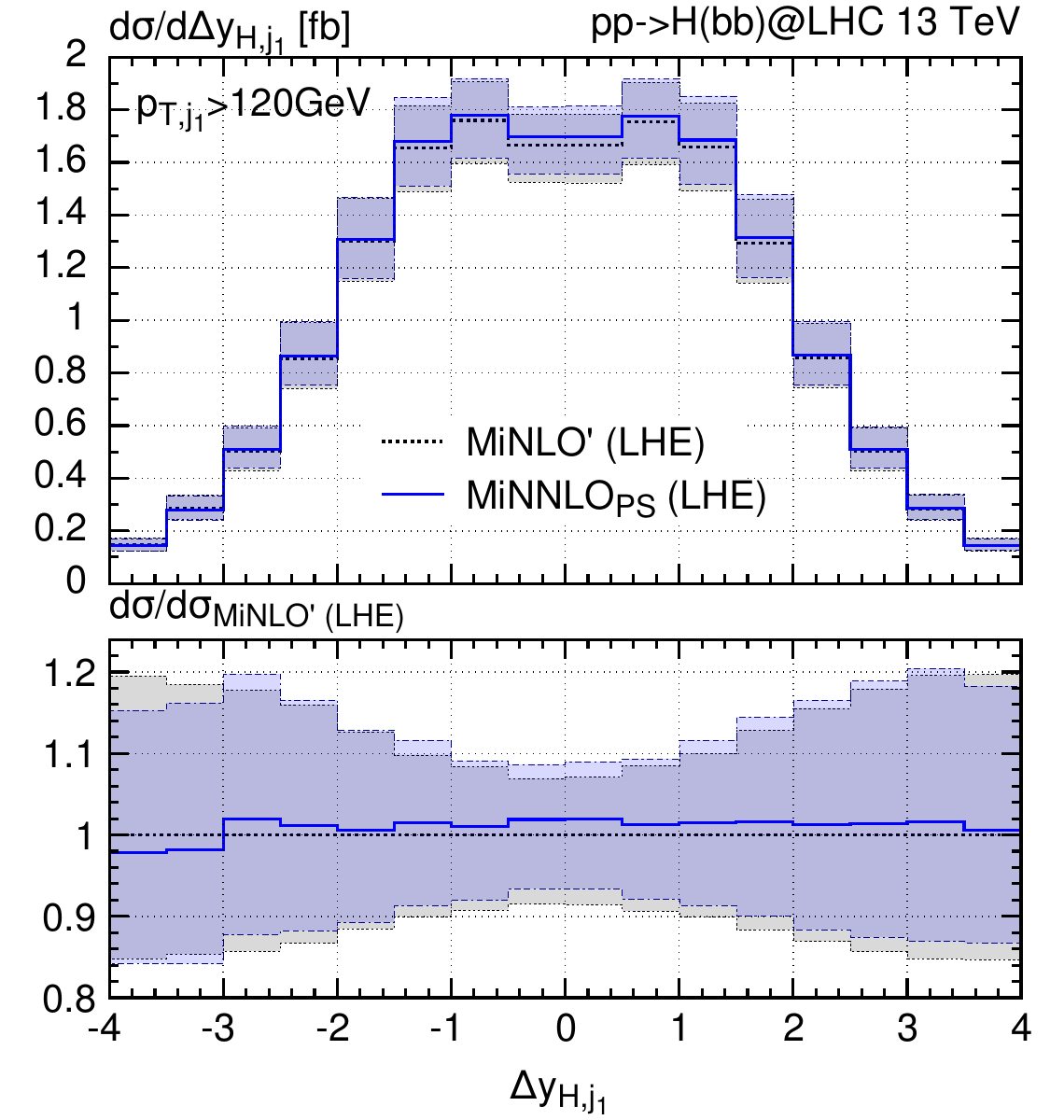}
\end{tabular}
\vspace*{1ex}
\caption{The left plots show the rapidity distribution of the leading jet ($y_{j_1}$) and the right plots show the rapidity difference between the Higgs boson and the leading 
jet ($\Delta y_{H,j_1}$). In the upper (lower) plots a transverse-momentum cut of $p_{T,j_1}>30$\,GeV ($p_{T,j_1}>120$\,GeV) is imposed on the leading jet.\label{fig:jetdist}}
\end{center}
\end{figure}

We continue by analyzing differential distributions. In \fig{fig:yHNNLO}, we compare
 our \minnlo{} predictions for the distribution in the  rapidity  of the Higgs boson ($y_H$) against the NNLO results~\cite{Mondini:2021nck}.\footnote{We thank the authors of \citere{Mondini:2021nck} for 
 providing us with the relevant results.} 
To facilitate this comparison, we adapted the input settings from our default ones, which were outlined before, to the ones employed in the NNLO computation of \citere{Mondini:2021nck}. Specifically, we use the NNLO set of the CT14 PDFs \cite{Dulat:2015mca} and we set the scales to $\muR=\KR m_H$ and $\muF=\KF m_H/4$. We find
a good agreement, both in terms of normalization and in terms of shape, 
between the two central predictions, with any difference being fully covered by the scale uncertainties. When looking at the scale-uncertainty bands, we notice some differences: the bands of \minnlo{} result are more symmetric and they are slightly smaller than the NNLO ones, where the central prediction lies at the upper edge of the prediction for central
rapidities.
Apart from that, we have checked that impact of the parton shower on the $y_H$ distribution is very moderate. 
We note that for the results presented in the remainder of this paper we will employ 
again our default input settings outlined at the beginning of this section.

In \fig{fig:jetdist}, we consider distributions that require the presence of at least one jet and compare \minnlo{} (blue, solid curve) with \minlo{} (black, dotted curve) predictions. 
The left plots show the rapidity distribution of the leading jet ($y_{j_1}$) and the right plots show the rapidity difference between the Higgs boson and the leading 
jet ($\Delta y_{H,j_1}$). In the upper (lower) plots a transverse-momentum cut of $p_{T,j_1}>30$\,GeV ($p_{T,j_1}>120$\,GeV) is imposed on the leading jet. For these distributions \minnlo{} and \minlo{} predictions
are both formally NLO QCD accurate. The purpose of this comparison is to validate that the NNLO corrections added by \minnlo{} do not alter significantly the \minlo{} result.
Indeed, we observe that the \minlo{} and \minnlo{} results have very similar shapes and that they are fully consistent within the quoted scale uncertainties. Moreover, the harder the required jet is, i.e. by
increasing $p_{T,j_1}$, the more similar \minlo{} and \minnlo{} predictions become.
The distribution in the rapidity difference between the Higgs and
the leading jet is, as expected, peaked around zero for
$p_{T,j_1}>30$\,GeV. This is due to the fact that the jet and the
Higgs boson are approximately balanced in transverse momentum, which
typically leads to the rapidity difference being centered around zero.
By contrast, when the leading jet is boosted (i.e.\ for
$p_{T,j_1}>120$\,GeV), the Higgs boson and jet tend to be slightly
farther apart in rapidity and we observe a dip in the $\Delta y_{H,j_1}$ 
distributions at central rapitidies.

\begin{figure}[t]
\begin{center}
\begin{tabular}{cc}
\includegraphics[width=.45\textwidth, page=1]{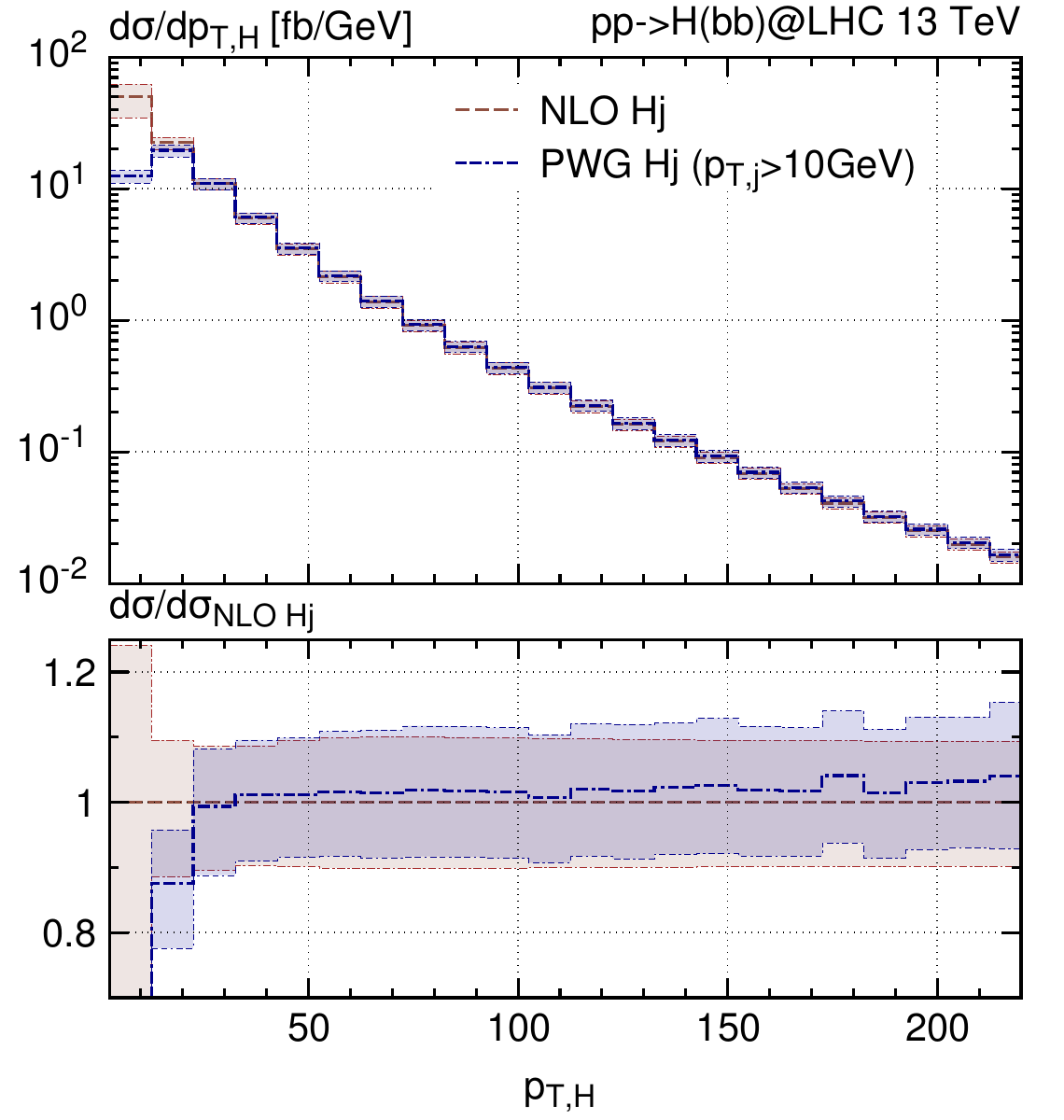}
&
\hspace{-0.8cm}
\includegraphics[width=.45\textwidth, page=1]{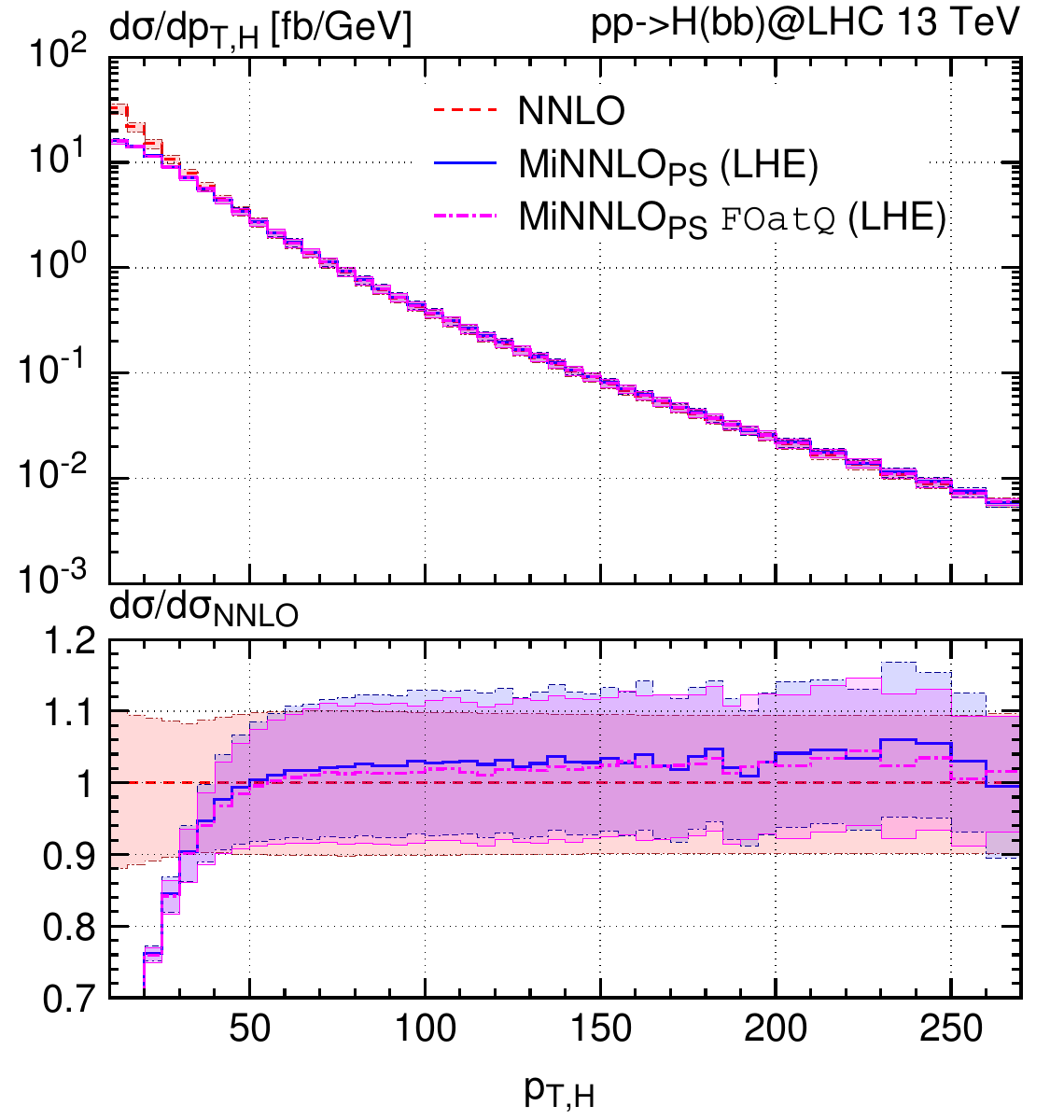}
\end{tabular}
\vspace*{1ex}
\caption{The left plot shows our \POWHEG{} implementation for $H$+jet production (dark-blue, double-dash-dotted curve) and a fixed-order calculation for $H$+jet production obtained from \citere{Harlander:2010cz} (brown, long-dashed curve) requiring $p_{T,j}>10$\,GeV. The right plot shows the analytic $p_{T,H}$ spectrum 
up to $\alpha_s^2$ from \citere{Harlander:2014hya} (red, dashed) and our \minnlo{} prediction with (magenta, dash-dotted curve) and without the {\tt FOatQ\,1} scale setting (blue, solid curve). \label{fig:largept}}
\end{center}
\end{figure}

Next, we consider the transverse-momentum spectrum of the Higgs boson ($p_{T,H}$), focusing on the large-$p_{T,H}$ region.
Here we validate our \minnlo{} generator, which formally is again only NLO accurate in QCD for $p_{T,H}\gtrsim m_H$,
against appropriate fixed-order calculations. In \fig{fig:largept} (left) we compare our \POWHEG{} implementation for $H$+jet production (dark-blue, double-dash-dotted curve)
with a fixed-order calculation for $H$+jet production obtained from \citere{Harlander:2010cz} (brown, long-dashed curve) requiring $p_{T,j}>10$\,GeV, while \fig{fig:largept} (right) shows the analytic $p_{T,H}$ spectrum 
up to $\alpha_s^2$ from \citere{Harlander:2014hya} (red, dashed) and our \minnlo{} prediction with (magenta, dash-dotted curve) and without the {\tt FOatQ 1} scale setting (blue, solid curve).
As expected, we find full agreement between our predictions and these fixed-order predictions in the regime of large $p_{T,H}$. By contrast, the fixed-order calculations
yield a divergent cross section for $p_{T,H}\to 0$, while the \minnlo{} prediction is finite in this region. We will study the low $p_{T,H}$ region in more detail at the end of this section.

\begin{figure}[t!]
\begin{center}
\begin{tabular}{cc}
\includegraphics[width=.45\textwidth]{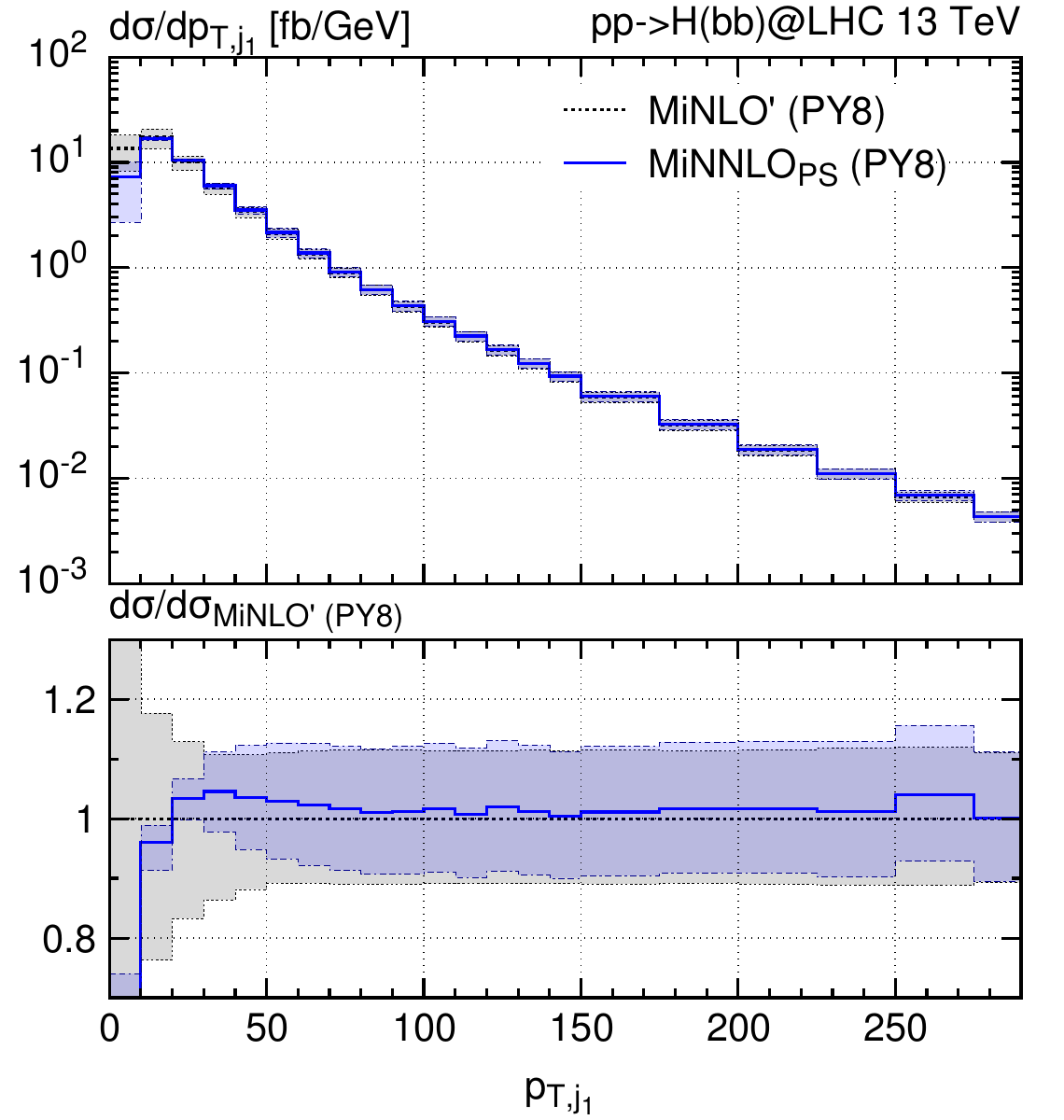}
&
\hspace{-0.8cm}
\includegraphics[width=.45\textwidth]{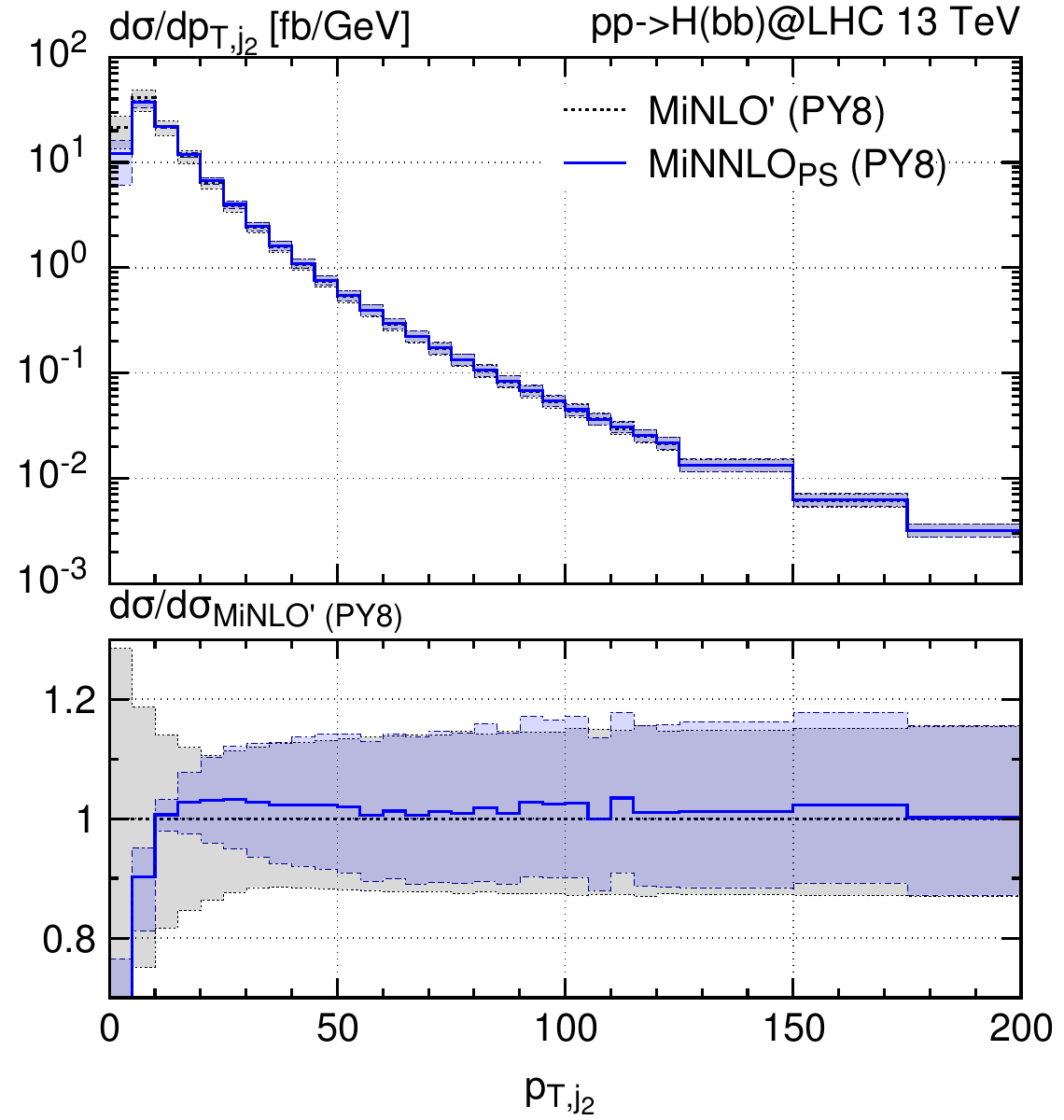}\\
\includegraphics[width=.45\textwidth]{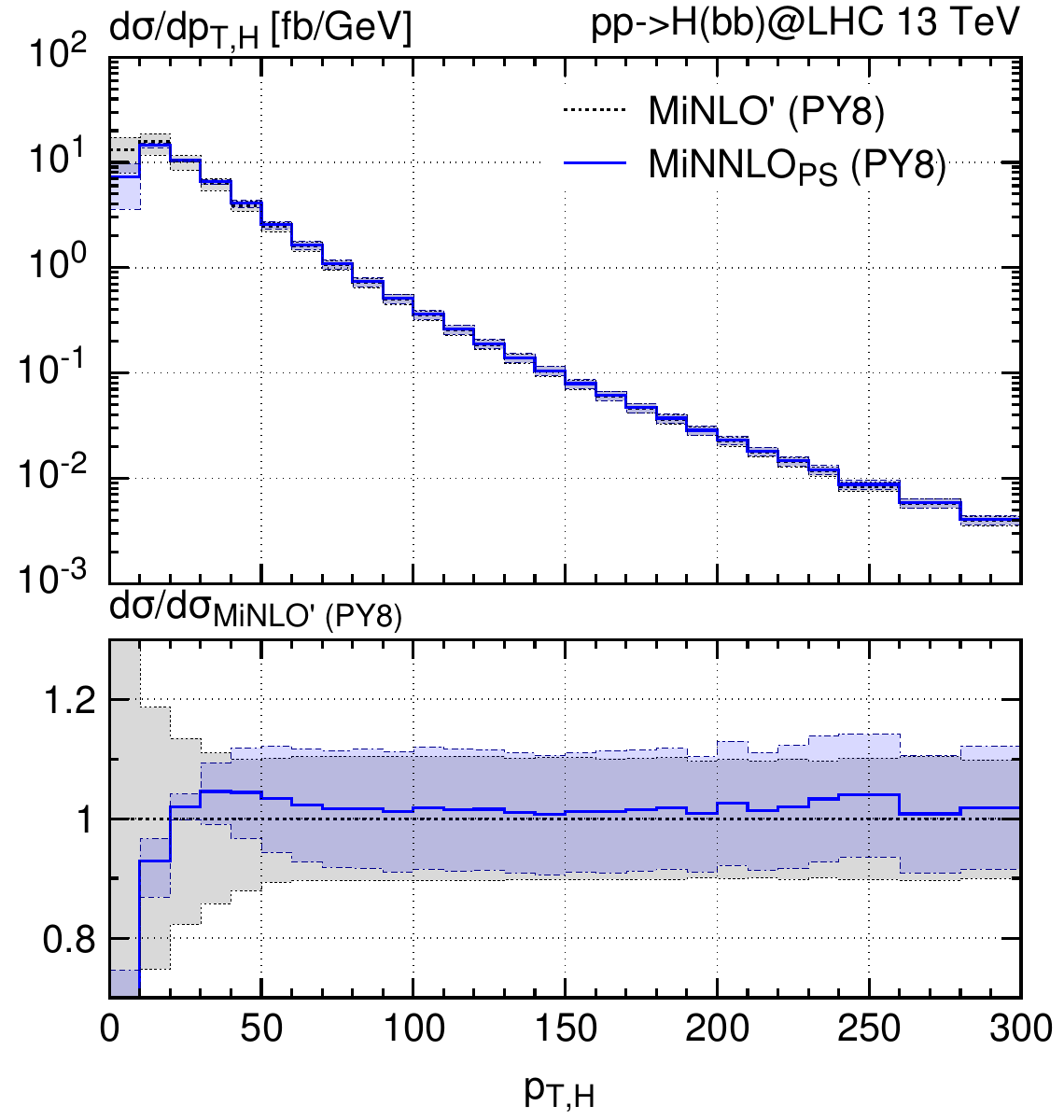}
&
\hspace{-0.8cm}
\includegraphics[width=.45\textwidth]{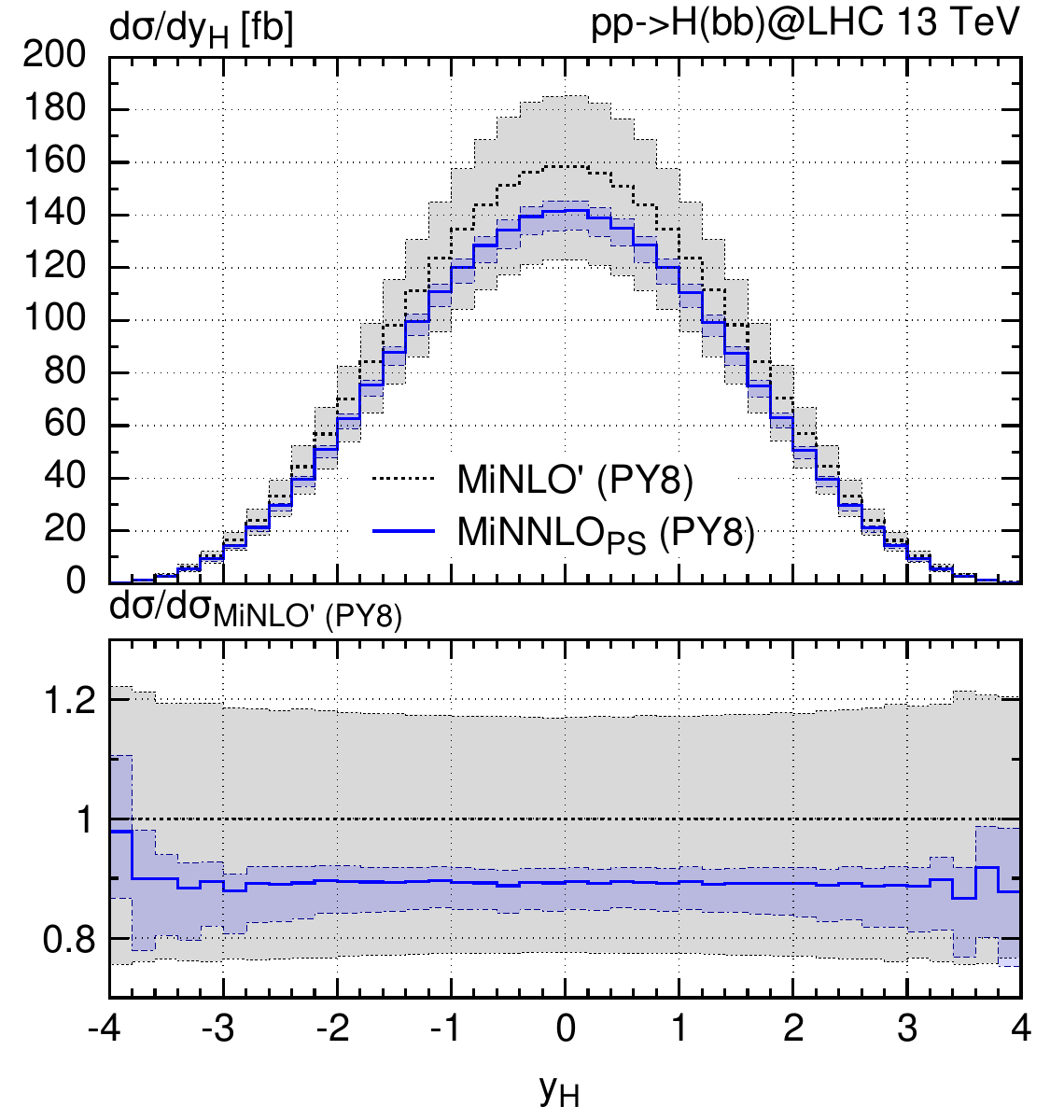}
\end{tabular}
\vspace*{1ex}
\caption{Comparison of \minlo{} (black, dotted) with \minnlo{} predictions (blue, solid) for the transverse momentum of the leading jet ($p_{T,j_1}$), the 
subleading jet ($p_{T,j_2}$), and the Higgs boson ($p_{T,H}$) as well as the rapidity distribution of the Higgs boson ($y_H$). \label{fig:minlovsminnlo}}
\end{center}
\end{figure}

We now turn to the comparison of \minlo{} with \minnlo{} predictions at the  fully differential level in \fig{fig:minlovsminnlo}, which allows us to assess
this size of the \minnlo{} corrections on top of \minlo{}. 
Specifically, the transverse momentum distributions of the leading jet ($p_{T,j_1}$),
subleading jet ($p_{T,j_2}$) and the Higgs boson ($p_{T,H}$) as well as the rapidity distribution of the Higgs boson ($y_H$) are shown.
We observe that the \minnlo{} corrections have a significant impact at small transverse momenta, dampening the spectrum significantly more in this region. Moreover,
scale uncertainties are substantially reduced in that region. On the contrary, at large transverse momenta the two predictions essentially coincide 
with each other, given that they are of the same formal accuracy in that region. We stress that these results are obtained using the correlated scale variation ($\KRy=\KR$). However, we also isolated the effects of the Yukawa scale variation fixing $\KR=1$ and varying just $\KRy$. In the low $p_{T,H}$ region, we obtained more symmetric uncertainties reflecting the previous observation on the total cross-section, while we notice that in the tail of the transverse momentum \
spectrum the scale variation is symmetric for both the settings, with a bigger uncertainty band for the correlated case $\KRy=\KR$. Finally, looking at the $y_H$ distribution, we observe a rather constant
and flat negative correction of about $-12\%$ due to \minnlo{} and a significant reduction of the scale uncertainties.

\begin{figure}[t!]
\begin{center}
\begin{tabular}{cc}
\includegraphics[width=.45\textwidth, page=1]{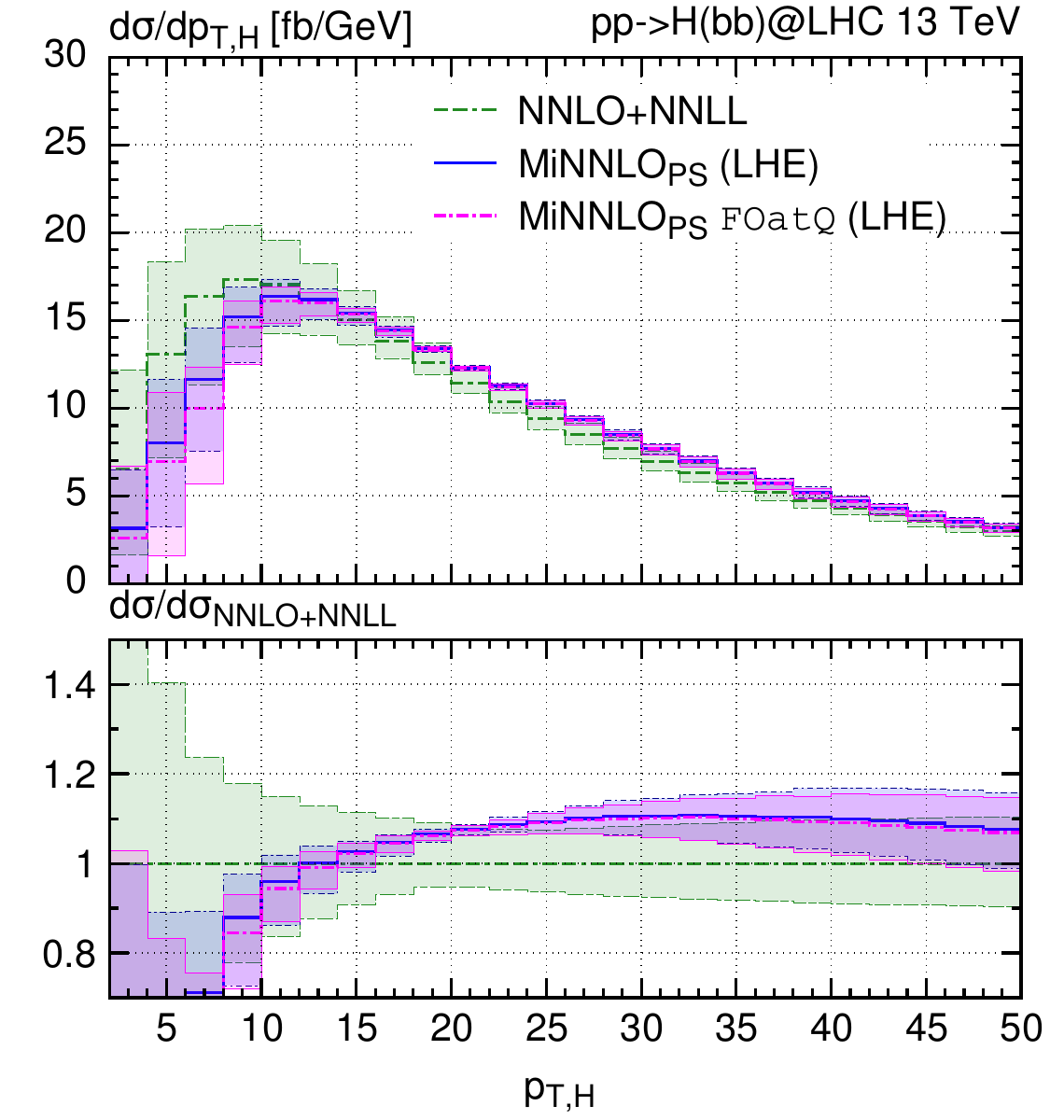}
&
\hspace{-0.8cm}
\includegraphics[width=.45\textwidth, page=1]{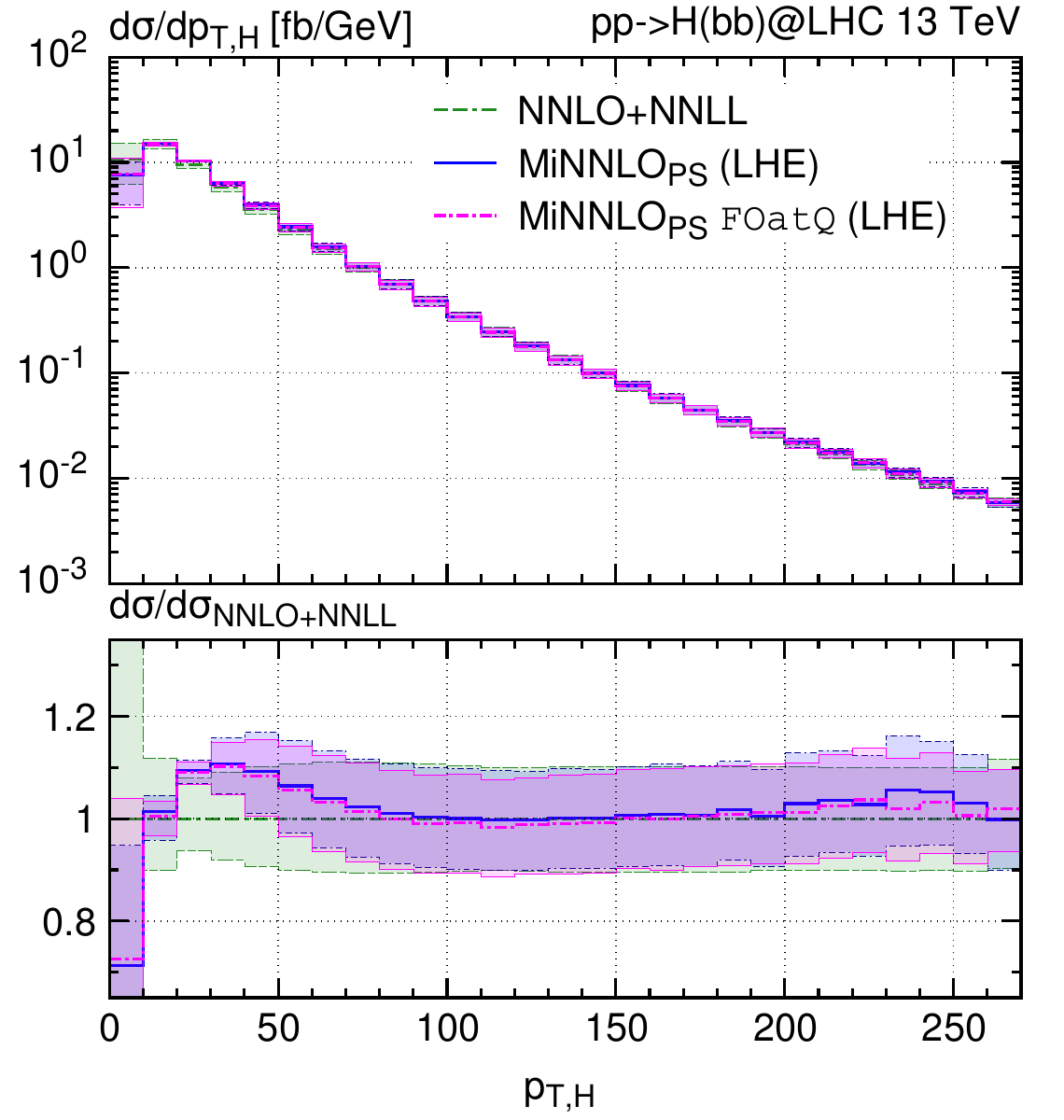}\\
\includegraphics[width=.45\textwidth, page=1]{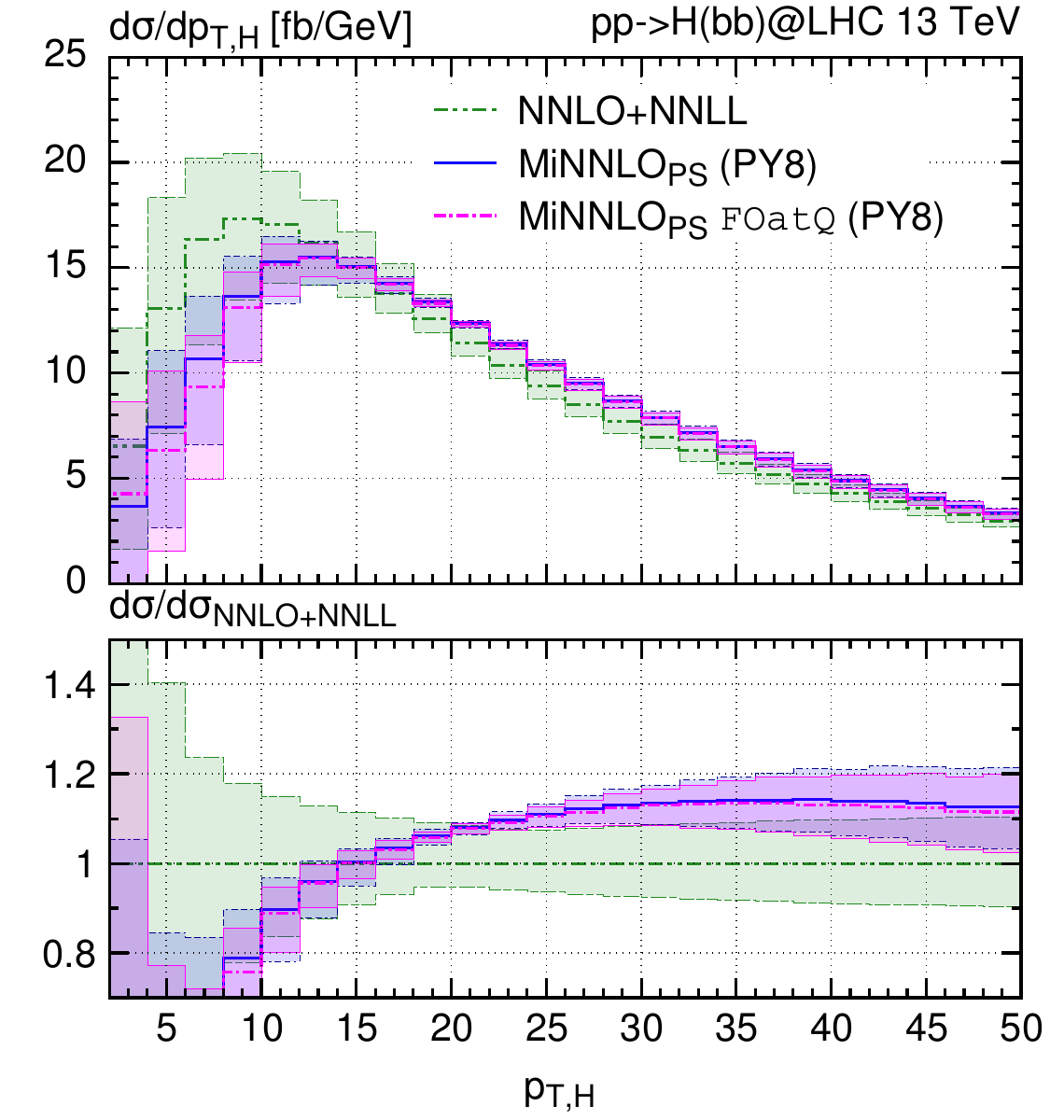}
&
\hspace{-0.8cm}
\includegraphics[width=.45\textwidth, page=1]{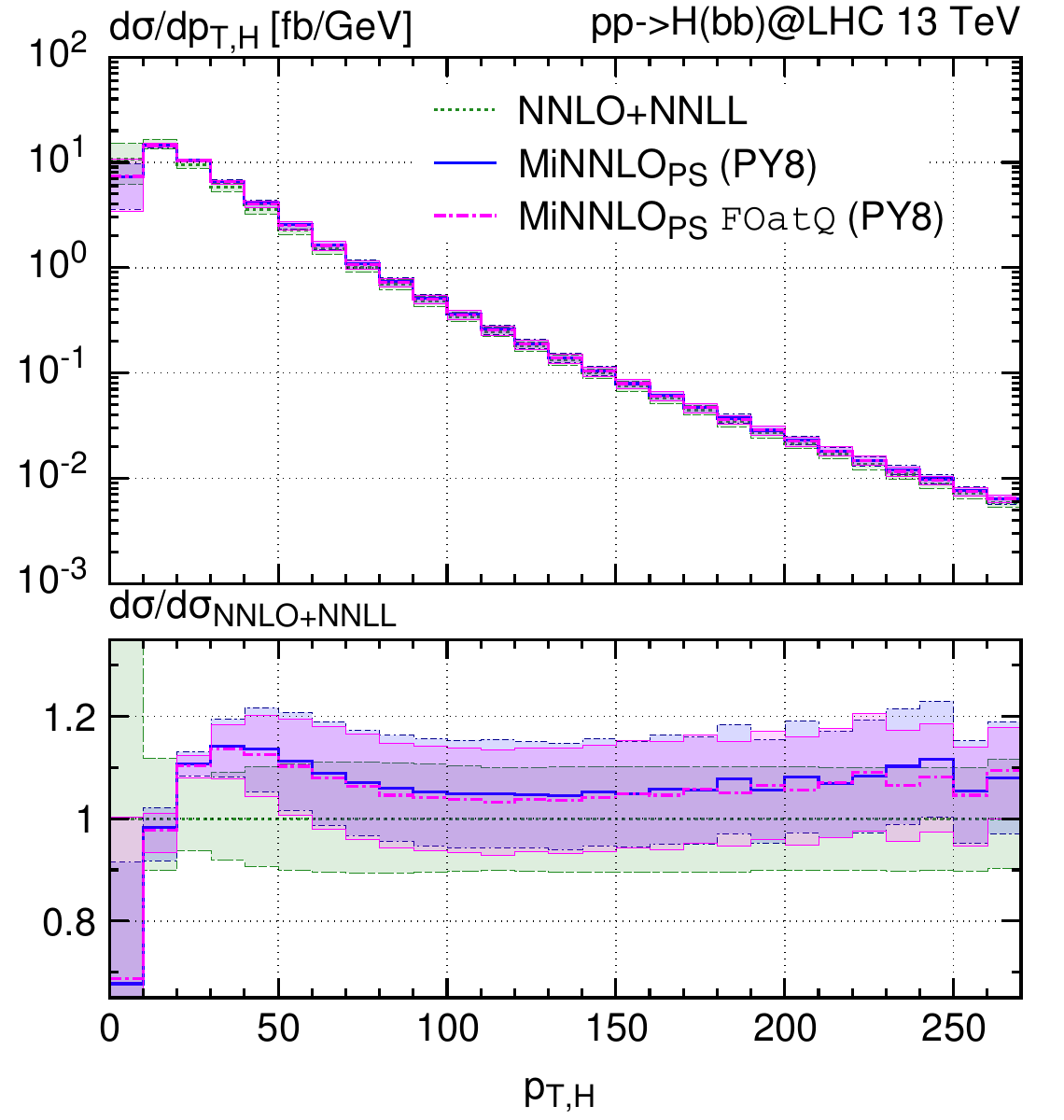}
\end{tabular}
\vspace*{1ex}
\caption{Predictions for the $p_{T,H}$ spectrum from our \minnlo{} generator compared to the analytic resummation at NNLO+NNLL of \citere{Harlander:2014hya} (green, double-dash-dotted curve)\label{fig:pthresum}.}
\end{center}
\end{figure}

We end this section by comparing the predictions for the $p_{T,H}$ spectrum from our \minnlo{} generator against
the analytic resummation at NNLO+NNLL of \citere{Harlander:2014hya} (green, double-dash-dotted curve) 
in \fig{fig:pthresum}. The left plots show this comparison zoomed into the low $p_{T,H}$ region, while the 
right ones include the $p_{T,H}$ spectrum up to 270\,GeV. The upper plots include \minnlo{} results
at the LHE level and the lower ones after showering with \PYTHIA{8}.
We find that again \minnlo{} results with and without the {\tt FOatQ 1} setting are very close. By contrast, 
the NNLO+NNLL prediction is softer than the \minnlo{} ones in the low $p_{T,H}$ region, where peak
of the spectrum is shifted towards higher values for \minnlo{}. At high $p_{T,H}$, where the predictions 
have the same accuracy, they coincide again, 
when considering the LHE-level \minnlo{} results.
Including the parton showering effects from \PYTHIA{8} tends to slightly worsen the agreement 
at small $p_{T,H}$, moving \minnlo{} away from the NNLO+NNLL scale band.
Considering the fact that the \minnlo{} predictions feature smaller scale uncertainties 
at small $p_{T,H}$ than the more accurate NNLO+NNLL prediction, it is clear that the \minnlo{} scale 
band does not reflect the actual size of uncertainties at small $p_{T,H}$, which would require additional 
variations within the shower settings. Since the assessment of the \PYTHIA{8} uncertainties is an 
issue intrinsic to the shower itself and not related to the NNLO+PS matching, we refrain from further studying 
these aspects here.
We note, however, that in the small $p_{T,H}$ region a massless approximation misses
potentially relevant mass effects, such that a combination with a massive 
4FS calculation might be needed,
which is beyond the scope of this paper and left for future work.
The shower also induces some effect at large $p_{T,H}$ moving the \minnlo{} prediction up by about 10\%, 
which is however well within the given scale-uncertainty bands.

\section{Summary}
We have presented a first NNLO+PS calculation of the $b\bar{b}\to H$ production process at the LHC.
To this end, we have revised the \minnlo{} method to account for an overall Yukawa coupling 
that is renormalized in the $\overline{\rm MS}$ scheme. Moreover, we have implemented an alternative 
approach to set the renormalization and factorization scales of the NLO 
colour-singlet plus jet calculation within \minnlo{}.

We have performed an extensive validation of our \minnlo{}
predictions, by comparing against inclusive NNLO QCD fixed-order results
for the total cross section as well as the rapidity distribution,
against lower-order \minlo{} results for
jet-related quantities, and by comparing to NLO QCD accurate
calculations at large values of the Higgs transverse momentum. In all
cases, we found consistency of our \minnlo{} results in the relevant
kinematical regions. At last, we considered the Higgs
transverse-momentum spectrum and showed a comparison of \minnlo{}
against an analytically resummed calculation at
NNLO+NNLL. Uncertainties of the \minnlo{} predictions seem to be
underestimated at small-$p_{T,H}$, given that they are smaller than
the ones of the more accurate NNLO+NNLL prediction and that the two
results barely agree within the quoted uncertainties. The quoted
uncertainties for \minnlo{} predictions however do not include
uncertainties related to the parton shower. The latter requires a
dedicated study which we leave to the interested reader.

We reckon that this \minnlo{} generator could be useful not only in
the context of direct searches for $b\bar{b}H$ final states, in
particular in some BSM context, but also for $HH$ measurements, where
the $b\bar{b}H$ process constitutes a major irreducible background.
Moreover, the present 5FS calculation of this process shall be
considered as a first step to a full NNLO+PS description of
$b\bar{b}H$ production, with the prospect of a 4FS $b\bar{b}H$
calculation with massive bottom quarks and eventually a full 4FS--5FS
combination at NNLO+PS accuracy, which we leave for future work.
Such combination could be particularly relevant in the small $p_{T,H}$ region 
where mass effects of the bottom quarks are potentially relevant.

\noindent {\bf Acknowledgements.}
We would like to thank Ciaran Williams for providing us with numbers for the 
NNLO rapidity distribution of the Higgs boson in $b\bar{b}\to H$ production.
We have used the Max Planck Computing and Data Facility (MPCDF) in Garching to carry 
out all simulations presented here.

\setlength{\bibsep}{3.1pt}
\renewcommand{\em}{}
\normalem
\bibliographystyle{JHEP}
\bibliography{MiNNLO_bbH}

\providecommand{\href}[2]{#2}\begingroup\raggedright\begin{thebibliography}{10}

\bibitem{ATLAS:2012yve}
{\scshape ATLAS} collaboration, \emph{{Observation of a new particle in the
  search for the Standard Model Higgs boson with the ATLAS detector at the
  LHC}}, \href{https://doi.org/10.1016/j.physletb.2012.08.020}{\emph{Phys.
  Lett. B} {\bfseries 716} (2012) 1}
  [\href{https://arxiv.org/abs/1207.7214}{{\ttfamily 1207.7214}}].

\bibitem{CMS:2012qbp}
{\scshape CMS} collaboration, \emph{{Observation of a New Boson at a Mass of
  125 GeV with the CMS Experiment at the LHC}},
  \href{https://doi.org/10.1016/j.physletb.2012.08.021}{\emph{Phys. Lett. B}
  {\bfseries 716} (2012) 30} [\href{https://arxiv.org/abs/1207.7235}{{\ttfamily
  1207.7235}}].

\bibitem{ATLAS:2022vkf}
{\scshape ATLAS} collaboration, \emph{{A detailed map of Higgs boson
  interactions by the ATLAS experiment ten years after the discovery}},
  \href{https://doi.org/10.1038/s41586-022-04893-w}{\emph{Nature} {\bfseries
  607} (2022) 52} [\href{https://arxiv.org/abs/2207.00092}{{\ttfamily
  2207.00092}}].

\bibitem{CMS:2022dwd}
{\scshape CMS} collaboration, \emph{{A portrait of the Higgs boson by the CMS
  experiment ten years after the discovery}},
  \href{https://doi.org/10.1038/s41586-022-04892-x}{\emph{Nature} {\bfseries
  607} (2022) 60} [\href{https://arxiv.org/abs/2207.00043}{{\ttfamily
  2207.00043}}].

\bibitem{Pagani:2020rsg}
D.~Pagani, H.-S.~Shao and M.~Zaro, \emph{{RIP $ Hb\overline{b} $: how other
  Higgs production modes conspire to kill a rare signal at the LHC}},
  \href{https://doi.org/10.1007/JHEP11(2020)036}{\emph{JHEP} {\bfseries 11}
  (2020) 036} [\href{https://arxiv.org/abs/2005.10277}{{\ttfamily
  2005.10277}}].

\bibitem{DiMicco:2019ngk}
J.~Alison et~al., \emph{{Higgs boson potential at colliders: Status and
  perspectives}}, \href{https://doi.org/10.1016/j.revip.2020.100045}{\emph{Rev.
  Phys.} {\bfseries 5} (2020) 100045}
  [\href{https://arxiv.org/abs/1910.00012}{{\ttfamily 1910.00012}}].

\bibitem{ATLAS:2022faz}
{\scshape ATLAS} collaboration, \emph{{HL-LHC prospects for the measurement of
  Higgs boson pair production in the $b\bar{b}b\bar{b}$ final state and
  combination with the $b\bar{b}\gamma\gamma$ and $b\bar{b}\tau^+\tau^-$ final
  states at the ATLAS experiment}}, .

\bibitem{Dicus:1998hs}
D.~Dicus, T.~Stelzer, Z.~Sullivan and S.~Willenbrock, \emph{{Higgs boson
  production in association with bottom quarks at next-to-leading order}},
  \href{https://doi.org/10.1103/PhysRevD.59.094016}{\emph{Phys.Rev.} {\bfseries
  D59} (1999) 094016} [\href{https://arxiv.org/abs/hep-ph/9811492}{{\ttfamily
  hep-ph/9811492}}].

\bibitem{Balazs:1998sb}
C.~Balazs, H.-J.~He and C.~Yuan, \emph{{QCD corrections to scalar production
  via heavy quark fusion at hadron colliders}},
  \href{https://doi.org/10.1103/PhysRevD.60.114001}{\emph{Phys.Rev.} {\bfseries
  D60} (1999) 114001} [\href{https://arxiv.org/abs/hep-ph/9812263}{{\ttfamily
  hep-ph/9812263}}].

\bibitem{Harlander:2003ai}
R.V.~Harlander and W.B.~Kilgore, \emph{{Higgs boson production in bottom quark
  fusion at next-to-next-to leading order}},
  \href{https://doi.org/10.1103/PhysRevD.68.013001}{\emph{Phys.Rev.} {\bfseries
  D68} (2003) 013001} [\href{https://arxiv.org/abs/hep-ph/0304035}{{\ttfamily
  hep-ph/0304035}}].

\bibitem{Campbell:2002zm}
J.M.~Campbell, R.K.~Ellis, F.~Maltoni and S.~Willenbrock, \emph{{Higgs-Boson
  production in association with a single bottom quark}},
  \href{https://doi.org/10.1103/PhysRevD.67.095002}{\emph{Phys.Rev.} {\bfseries
  D67} (2003) 095002} [\href{https://arxiv.org/abs/hep-ph/0204093}{{\ttfamily
  hep-ph/0204093}}].

\bibitem{Harlander:2010cz}
R.V.~Harlander, K.J.~Ozeren and M.~Wiesemann, \emph{{Higgs plus jet production
  in bottom quark annihilation at next-to-leading order}},
  \href{https://doi.org/10.1016/j.physletb.2010.08.038}{\emph{Phys. Lett.}
  {\bfseries B693} (2010) 269}
  [\href{https://arxiv.org/abs/1007.5411}{{\ttfamily 1007.5411}}].

\bibitem{Ozeren:2010qp}
K.J.~Ozeren, \emph{{Analytic Results for Higgs Production in Bottom Fusion}},
  \href{https://doi.org/10.1007/JHEP11(2010)084}{\emph{JHEP} {\bfseries 1011}
  (2010) 084} [\href{https://arxiv.org/abs/1010.2977}{{\ttfamily 1010.2977}}].

\bibitem{Harlander:2011fx}
R.~Harlander and M.~Wiesemann, \emph{{Jet-veto in bottom-quark induced Higgs
  production at next-to-next-to-leading order}},
  \href{https://doi.org/10.1007/JHEP04(2012)066}{\emph{JHEP} {\bfseries 1204}
  (2012) 066} [\href{https://arxiv.org/abs/1111.2182}{{\ttfamily 1111.2182}}].

\bibitem{Buehler:2012cu}
S.~Buhler, F.~Herzog, A.~Lazopoulos and R.~Muller, \emph{{The fully
  differential hadronic production of a Higgs boson via bottom quark fusion at
  NNLO}}, \href{https://doi.org/10.1007/JHEP07(2012)115}{\emph{JHEP} {\bfseries
  1207} (2012) 115} [\href{https://arxiv.org/abs/1204.4415}{{\ttfamily
  1204.4415}}].

\bibitem{Belyaev:2005bs}
A.~Belyaev, P.M.~Nadolsky and C.-P.~Yuan, \emph{{Transverse momentum
  resummation for Higgs boson produced via b anti-b fusion at hadron
  colliders}}, \href{https://doi.org/10.1088/1126-6708/2006/04/004}{\emph{JHEP}
  {\bfseries 0604} (2006) 004}
  [\href{https://arxiv.org/abs/hep-ph/0509100}{{\ttfamily hep-ph/0509100}}].

\bibitem{Harlander:2014hya}
R.V.~Harlander, A.~Tripathi and M.~Wiesemann, \emph{{Higgs production in bottom
  quark annihilation: Transverse momentum distribution at NNLO$+$NNLL}},
  \href{https://doi.org/10.1103/PhysRevD.90.015017}{\emph{Phys. Rev.}
  {\bfseries D90} (2014) 015017}
  [\href{https://arxiv.org/abs/1403.7196}{{\ttfamily 1403.7196}}].

\bibitem{Ahmed:2014pka}
T.~Ahmed, M.~Mahakhud, P.~Mathews, N.~Rana and V.~Ravindran, \emph{{Two-loop
  QCD corrections to Higgs $\to b+\overline{b}+g$ amplitude}},
  \href{https://doi.org/10.1007/JHEP08(2014)075}{\emph{JHEP} {\bfseries 1408}
  (2014) 075} [\href{https://arxiv.org/abs/1405.2324}{{\ttfamily 1405.2324}}].

\bibitem{Gehrmann:2014vha}
T.~Gehrmann and D.~Kara, \emph{{The $Hb\bar{b}$ form factor to three loops in
  QCD}}, \href{https://doi.org/10.1007/JHEP09(2014)174}{\emph{JHEP} {\bfseries
  09} (2014) 174} [\href{https://arxiv.org/abs/1407.8114}{{\ttfamily
  1407.8114}}].

\bibitem{Duhr:2019kwi}
C.~Duhr, F.~Dulat and B.~Mistlberger, \emph{{Higgs Boson Production in
  Bottom-Quark Fusion to Third Order in the Strong Coupling}},
  \href{https://doi.org/10.1103/PhysRevLett.125.051804}{\emph{Phys. Rev. Lett.}
  {\bfseries 125} (2020) 051804}
  [\href{https://arxiv.org/abs/1904.09990}{{\ttfamily 1904.09990}}].

\bibitem{Mondini:2021nck}
R.~Mondini and C.~Williams, \emph{{Bottom-induced contributions to Higgs plus
  jet at next-to-next-to-leading order}},
  \href{https://doi.org/10.1007/JHEP05(2021)045}{\emph{JHEP} {\bfseries 05}
  (2021) 045} [\href{https://arxiv.org/abs/2102.05487}{{\ttfamily
  2102.05487}}].

\bibitem{Wiesemann:2014ioa}
M.~Wiesemann, R.~Frederix, S.~Frixione, V.~Hirschi, F.~Maltoni and
  P.~Torrielli, \emph{{Higgs production in association with bottom quarks}},
  \href{https://doi.org/10.1007/JHEP02(2015)132}{\emph{JHEP} {\bfseries 02}
  (2015) 132} [\href{https://arxiv.org/abs/1409.5301}{{\ttfamily 1409.5301}}].

\bibitem{Krauss:2016orf}
F.~Krauss, D.~Napoletano and S.~Schumann, \emph{{Simulating $b$-associated
  production of $Z$ and Higgs bosons with the SHERPA event generator}},
  \href{https://doi.org/10.1103/PhysRevD.95.036012}{\emph{Phys. Rev.}
  {\bfseries D95} (2017) 036012}
  [\href{https://arxiv.org/abs/1612.04640}{{\ttfamily 1612.04640}}].

\bibitem{Ajjath:2019ixh}
A.H.~Ajjath, P.~Banerjee, A.~Chakraborty, P.K.~Dhani, P.~Mukherjee, N.~Rana
  et~al., \emph{{NNLO QCD$\oplus$QED corrections to Higgs production in bottom
  quark annihilation}},
  \href{https://doi.org/10.1103/PhysRevD.100.114016}{\emph{Phys. Rev. D}
  {\bfseries 100} (2019) 114016}
  [\href{https://arxiv.org/abs/1906.09028}{{\ttfamily 1906.09028}}].

\bibitem{Ajjath:2019neu}
A.H.~Ajjath, A.~Chakraborty, G.~Das, P.~Mukherjee and V.~Ravindran,
  \emph{{Resummed prediction for Higgs boson production through b$
  \overline{\mathrm{b}} $ annihilation at N$^{3}$LL}},
  \href{https://doi.org/10.1007/JHEP11(2019)006}{\emph{JHEP} {\bfseries 11}
  (2019) 006} [\href{https://arxiv.org/abs/1905.03771}{{\ttfamily
  1905.03771}}].

\bibitem{Forte:2019hjc}
S.~Forte, T.~Giani and D.~Napoletano, \emph{{Fitting the b-quark PDF as a
  massive-b scheme: Higgs production in bottom fusion}},
  \href{https://doi.org/10.1140/epjc/s10052-019-7119-3}{\emph{Eur. Phys. J. C}
  {\bfseries 79} (2019) 609}
  [\href{https://arxiv.org/abs/1905.02207}{{\ttfamily 1905.02207}}].

\bibitem{Badger:2021ega}
S.~Badger, H.B.~Hartanto, J.~Kry\'s and S.~Zoia, \emph{{Two-loop leading-colour
  QCD helicity amplitudes for Higgs boson production in association with a
  bottom-quark pair at the LHC}},
  \href{https://doi.org/10.1007/JHEP11(2021)012}{\emph{JHEP} {\bfseries 11}
  (2021) 012} [\href{https://arxiv.org/abs/2107.14733}{{\ttfamily
  2107.14733}}].

\bibitem{Dittmaier:2003ej}
S.~Dittmaier, M.~Kr{\"a}mer and M.~Spira, \emph{{Higgs radiation off bottom
  quarks at the Tevatron and the CERN LHC}},
  \href{https://doi.org/10.1103/PhysRevD.70.074010}{\emph{Phys.Rev.} {\bfseries
  D70} (2004) 074010} [\href{https://arxiv.org/abs/hep-ph/0309204}{{\ttfamily
  hep-ph/0309204}}].

\bibitem{Dawson:2003kb}
S.~Dawson, C.~Jackson, L.~Reina and D.~Wackeroth, \emph{{Exclusive Higgs boson
  production with bottom quarks at hadron colliders}},
  \href{https://doi.org/10.1103/PhysRevD.69.074027}{\emph{Phys.Rev.} {\bfseries
  D69} (2004) 074027} [\href{https://arxiv.org/abs/hep-ph/0311067}{{\ttfamily
  hep-ph/0311067}}].

\bibitem{Dawson:2005vi}
S.~Dawson, C.~Jackson, L.~Reina and D.~Wackeroth, \emph{{Higgs production in
  association with bottom quarks at hadron colliders}},
  \href{https://doi.org/10.1142/S0217732306019256}{\emph{Mod.Phys.Lett.}
  {\bfseries A21} (2006) 89}
  [\href{https://arxiv.org/abs/hep-ph/0508293}{{\ttfamily hep-ph/0508293}}].

\bibitem{Liu:2012qu}
N.~Liu, L.~Wu, P.W.~Wu and J.M.~Yang, \emph{{Complete one-loop effects of SUSY
  QCD in $b\bar{b}h$ production at the LHC under current experimental
  constraints}}, \href{https://doi.org/10.1007/JHEP01(2013)161}{\emph{JHEP}
  {\bfseries 1301} (2013) 161}
  [\href{https://arxiv.org/abs/1208.3413}{{\ttfamily 1208.3413}}].

\bibitem{Dittmaier:2014sva}
S.~Dittmaier, P.~H\"afliger, M.~Kr\"amer, M.~Spira and M.~Walser,
  \emph{{Neutral MSSM Higgs-boson production with heavy quarks: NLO
  supersymmetric QCD corrections}},
  \href{https://doi.org/10.1103/PhysRevD.90.035010}{\emph{Phys. Rev. D}
  {\bfseries 90} (2014) 035010}
  [\href{https://arxiv.org/abs/1406.5307}{{\ttfamily 1406.5307}}].

\bibitem{Zhang:2017mdz}
Y.~Zhang, \emph{{NLO electroweak effects on the Higgs boson production in
  association with a bottom quark pair at the LHC}},
  \href{https://doi.org/10.1103/PhysRevD.96.113009}{\emph{Phys. Rev.}
  {\bfseries D96} (2017) 113009}
  [\href{https://arxiv.org/abs/1708.08790}{{\ttfamily 1708.08790}}].

\bibitem{Jager:2015hka}
B.~Jager, L.~Reina and D.~Wackeroth, \emph{{Higgs boson production in
  association with b jets in the POWHEG BOX}},
  \href{https://doi.org/10.1103/PhysRevD.93.014030}{\emph{Phys. Rev.}
  {\bfseries D93} (2016) 014030}
  [\href{https://arxiv.org/abs/1509.05843}{{\ttfamily 1509.05843}}].

\bibitem{Deutschmann:2018avk}
N.~Deutschmann, F.~Maltoni, M.~Wiesemann and M.~Zaro, \emph{{Top-Yukawa
  contributions to bbH production at the LHC}},
  \href{https://doi.org/10.1007/JHEP07(2019)054}{\emph{JHEP} {\bfseries 07}
  (2019) 054} [\href{https://arxiv.org/abs/1808.01660}{{\ttfamily
  1808.01660}}].

\bibitem{Manzoni:2023qaf}
S.~Manzoni, E.~Mazzeo, J.~Mazzitelli, M.~Wiesemann and M.~Zaro, \emph{{Taming a
  leading theoretical uncertainty in HH measurements via accurate simulations
  for $ \textrm{b}\overline{\textrm{b}}\textrm{H} $ production}},
  \href{https://doi.org/10.1007/JHEP09(2023)179}{\emph{JHEP} {\bfseries 09}
  (2023) 179} [\href{https://arxiv.org/abs/2307.09992}{{\ttfamily
  2307.09992}}].

\bibitem{Maltoni:2012pa}
F.~Maltoni, G.~Ridolfi and M.~Ubiali, \emph{{b-initiated processes at the LHC:
  a reappraisal}}, \href{https://doi.org/10.1007/JHEP04(2013)095,
  10.1007/JHEP07(2012)022}{\emph{JHEP} {\bfseries 1207} (2012) 022}
  [\href{https://arxiv.org/abs/1203.6393}{{\ttfamily 1203.6393}}].

\bibitem{Lim:2016wjo}
M.~Lim, F.~Maltoni, G.~Ridolfi and M.~Ubiali, \emph{{Anatomy of double
  heavy-quark initiated processes}},
  \href{https://doi.org/10.1007/JHEP09(2016)132}{\emph{JHEP} {\bfseries 09}
  (2016) 132} [\href{https://arxiv.org/abs/1605.09411}{{\ttfamily
  1605.09411}}].

\bibitem{Forte:2015hba}
S.~Forte, D.~Napoletano and M.~Ubiali, \emph{{Higgs production in bottom-quark
  fusion in a matched scheme}},
  \href{https://doi.org/10.1016/j.physletb.2015.10.051}{\emph{Phys. Lett.}
  {\bfseries B751} (2015) 331}
  [\href{https://arxiv.org/abs/1508.01529}{{\ttfamily 1508.01529}}].

\bibitem{Forte:2016sja}
S.~Forte, D.~Napoletano and M.~Ubiali, \emph{{Higgs production in bottom-quark
  fusion: matching beyond leading order}},
  \href{https://doi.org/10.1016/j.physletb.2016.10.040}{\emph{Phys. Lett.}
  {\bfseries B763} (2016) 190}
  [\href{https://arxiv.org/abs/1607.00389}{{\ttfamily 1607.00389}}].

\bibitem{Bonvini:2015pxa}
M.~Bonvini, A.S.~Papanastasiou and F.J.~Tackmann, \emph{{Resummation and
  matching of b-quark mass effects in $ b\overline{b}H $ production}},
  \href{https://doi.org/10.1007/JHEP11(2015)196}{\emph{JHEP} {\bfseries 11}
  (2015) 196} [\href{https://arxiv.org/abs/1508.03288}{{\ttfamily
  1508.03288}}].

\bibitem{Bonvini:2016fgf}
M.~Bonvini, A.S.~Papanastasiou and F.J.~Tackmann, \emph{{Matched predictions
  for the $ b\overline{b}H $ cross section at the 13 TeV LHC}},
  \href{https://doi.org/10.1007/JHEP10(2016)053}{\emph{JHEP} {\bfseries 10}
  (2016) 053} [\href{https://arxiv.org/abs/1605.01733}{{\ttfamily
  1605.01733}}].

\bibitem{Duhr:2020kzd}
C.~Duhr, F.~Dulat, V.~Hirschi and B.~Mistlberger, \emph{{Higgs production in
  bottom quark fusion: matching the 4- and 5-flavour schemes to third order in
  the strong coupling}},
  \href{https://doi.org/10.1007/JHEP08(2020)017}{\emph{JHEP} {\bfseries 08}
  (2020) 017} [\href{https://arxiv.org/abs/2004.04752}{{\ttfamily
  2004.04752}}].

\bibitem{Monni:2019whf}
P.F.~Monni, P.~Nason, E.~Re, M.~Wiesemann and G.~Zanderighi,
  \emph{{MiNNLO$_{PS}$: a new method to match NNLO QCD to parton showers}},
  \href{https://doi.org/10.1007/JHEP05(2020)143}{\emph{JHEP} {\bfseries 05}
  (2020) 143} [\href{https://arxiv.org/abs/1908.06987}{{\ttfamily
  1908.06987}}].

\bibitem{Monni:2020nks}
P.F.~Monni, E.~Re and M.~Wiesemann, \emph{{MiNNLO$_{\text {PS}}$: optimizing
  $2\rightarrow 1$ hadronic processes}},
  \href{https://doi.org/10.1140/epjc/s10052-020-08658-5}{\emph{Eur. Phys. J. C}
  {\bfseries 80} (2020) 1075}
  [\href{https://arxiv.org/abs/2006.04133}{{\ttfamily 2006.04133}}].

\bibitem{Harlander:2012pb}
R.V.~Harlander, S.~Liebler and H.~Mantler, \emph{{SusHi: A program for the
  calculation of Higgs production in gluon fusion and bottom-quark annihilation
  in the Standard Model and the MSSM}},
  \href{https://doi.org/10.1016/j.cpc.2013.02.006}{\emph{Comput. Phys. Commun.}
  {\bfseries 184} (2013) 1605}
  [\href{https://arxiv.org/abs/1212.3249}{{\ttfamily 1212.3249}}].

\bibitem{Lombardi:2020wju}
D.~Lombardi, M.~Wiesemann and G.~Zanderighi, \emph{{Advancing M\i{}NNLO$_{PS}$
  to diboson processes: Z\ensuremath{\gamma} production at NNLO+PS}},
  \href{https://doi.org/10.1007/JHEP06(2021)095}{\emph{JHEP} {\bfseries 06}
  (2021) 095} [\href{https://arxiv.org/abs/2010.10478}{{\ttfamily
  2010.10478}}].

\bibitem{Lombardi:2021rvg}
D.~Lombardi, M.~Wiesemann and G.~Zanderighi, \emph{{WW production at NNLO+PS
  with MiNNLO$_{PS}$}},
  \href{https://doi.org/10.1007/JHEP11(2021)230}{\emph{JHEP} {\bfseries 11}
  (2021) 230} [\href{https://arxiv.org/abs/2103.12077}{{\ttfamily
  2103.12077}}].

\bibitem{Lombardi:2021wug}
D.~Lombardi, M.~Wiesemann and G.~Zanderighi, \emph{{Anomalous couplings in
  Z\ensuremath{\gamma} events at NNLO+PS and improving
  \ensuremath{\nu}\ensuremath{\nu}\textasciimacron{}\ensuremath{\gamma}
  backgrounds in dark-matter searches}},
  \href{https://doi.org/10.1016/j.physletb.2021.136846}{\emph{Phys. Lett. B}
  {\bfseries 824} (2022) 136846}
  [\href{https://arxiv.org/abs/2108.11315}{{\ttfamily 2108.11315}}].

\bibitem{Buonocore:2021fnj}
L.~Buonocore, G.~Koole, D.~Lombardi, L.~Rottoli, M.~Wiesemann and
  G.~Zanderighi, \emph{{ZZ production at nNNLO+PS with MiNNLO$_{PS}$}},
  \href{https://doi.org/10.1007/JHEP01(2022)072}{\emph{JHEP} {\bfseries 01}
  (2022) 072} [\href{https://arxiv.org/abs/2108.05337}{{\ttfamily
  2108.05337}}].

\bibitem{Zanoli:2021iyp}
S.~Zanoli, M.~Chiesa, E.~Re, M.~Wiesemann and G.~Zanderighi,
  \emph{{Next-to-next-to-leading order event generation for $VH$ production
  with $H\to b\bar{b}$ decay}},
  \href{https://arxiv.org/abs/2112.04168}{{\ttfamily 2112.04168}}.

\bibitem{Gavardi:2022ixt}
A.~Gavardi, C.~Oleari and E.~Re, \emph{{NNLO+PS Monte Carlo simulation of
  photon pair production with MiNNLO$_{PS}$}},
  \href{https://doi.org/10.1007/JHEP09(2022)061}{\emph{JHEP} {\bfseries 09}
  (2022) 061} [\href{https://arxiv.org/abs/2204.12602}{{\ttfamily
  2204.12602}}].

\bibitem{Haisch:2022nwz}
U.~Haisch, D.J.~Scott, M.~Wiesemann, G.~Zanderighi and S.~Zanoli, \emph{{NNLO
  event generation for $ pp\to Zh\to
  {\mathrm{\ell}}^{+}{\mathrm{\ell}}^{-}b\overline{b} $ production in the SM
  effective field theory}},
  \href{https://doi.org/10.1007/JHEP07(2022)054}{\emph{JHEP} {\bfseries 07}
  (2022) 054} [\href{https://arxiv.org/abs/2204.00663}{{\ttfamily
  2204.00663}}].

\bibitem{Lindert:2022qdd}
J.M.~Lindert, D.~Lombardi, M.~Wiesemann, G.~Zanderighi and S.~Zanoli,
  \emph{{W$^{±}$Z production at NNLO QCD and NLO EW matched to parton showers
  with MiNNLO$_{PS}$}},
  \href{https://doi.org/10.1007/JHEP11(2022)036}{\emph{JHEP} {\bfseries 11}
  (2022) 036} [\href{https://arxiv.org/abs/2208.12660}{{\ttfamily
  2208.12660}}].

\bibitem{Gauld:2023gtb}
R.~Gauld, U.~Haisch and L.~Schnell, \emph{{SMEFT at NNLO+PS: Vh production}},
  \href{https://doi.org/10.1007/JHEP01(2024)192}{\emph{JHEP} {\bfseries 01}
  (2024) 192} [\href{https://arxiv.org/abs/2311.06107}{{\ttfamily
  2311.06107}}].

\bibitem{Mazzitelli:2020jio}
J.~Mazzitelli, P.F.~Monni, P.~Nason, E.~Re, M.~Wiesemann and G.~Zanderighi,
  \emph{{Next-to-Next-to-Leading Order Event Generation for Top-Quark Pair
  Production}},
  \href{https://doi.org/10.1103/PhysRevLett.127.062001}{\emph{Phys. Rev. Lett.}
  {\bfseries 127} (2021) 062001}
  [\href{https://arxiv.org/abs/2012.14267}{{\ttfamily 2012.14267}}].

\bibitem{Mazzitelli:2021mmm}
J.~Mazzitelli, P.F.~Monni, P.~Nason, E.~Re, M.~Wiesemann and G.~Zanderighi,
  \emph{{Top-pair production at the LHC with MiNNLO$_{PS}$}},
  \href{https://doi.org/10.1007/JHEP04(2022)079}{\emph{JHEP} {\bfseries 04}
  (2022) 079} [\href{https://arxiv.org/abs/2112.12135}{{\ttfamily
  2112.12135}}].

\bibitem{Mazzitelli:2023znt}
J.~Mazzitelli, A.~Ratti, M.~Wiesemann and G.~Zanderighi, \emph{{B-hadron
  production at the LHC from bottom-quark pair production at NNLO+PS}},
  \href{https://doi.org/10.1016/j.physletb.2023.137991}{\emph{Phys. Lett. B}
  {\bfseries 843} (2023) 137991}
  [\href{https://arxiv.org/abs/2302.01645}{{\ttfamily 2302.01645}}].

\bibitem{Jezo:2015aia}
T.~Je\v{z}o and P.~Nason, \emph{{On the Treatment of Resonances in
  Next-to-Leading Order Calculations Matched to a Parton Shower}},
  \href{https://doi.org/10.1007/JHEP12(2015)065}{\emph{JHEP} {\bfseries 12}
  (2015) 065} [\href{https://arxiv.org/abs/1509.09071}{{\ttfamily
  1509.09071}}].

\bibitem{Nason:2004rx}
P.~Nason, \emph{{A New method for combining NLO QCD with shower Monte Carlo
  algorithms}},
  \href{https://doi.org/10.1088/1126-6708/2004/11/040}{\emph{JHEP} {\bfseries
  11} (2004) 040} [\href{https://arxiv.org/abs/hep-ph/0409146}{{\ttfamily
  hep-ph/0409146}}].

\bibitem{Alioli:2010xd}
S.~Alioli, P.~Nason, C.~Oleari and E.~Re, \emph{{A general framework for
  implementing NLO calculations in shower Monte Carlo programs: the POWHEG
  BOX}}, \href{https://doi.org/10.1007/JHEP06(2010)043}{\emph{JHEP} {\bfseries
  06} (2010) 043} [\href{https://arxiv.org/abs/1002.2581}{{\ttfamily
  1002.2581}}].

\bibitem{Frixione:2007vw}
S.~Frixione, P.~Nason and C.~Oleari, \emph{{Matching NLO QCD computations with
  Parton Shower simulations: the POWHEG method}},
  \href{https://doi.org/10.1088/1126-6708/2007/11/070}{\emph{JHEP} {\bfseries
  11} (2007) 070} [\href{https://arxiv.org/abs/0709.2092}{{\ttfamily
  0709.2092}}].

\bibitem{Cascioli:2011va}
F.~Cascioli, P.~Maierh\"ofer and S.~Pozzorini, \emph{{Scattering Amplitudes
  with Open Loops}},
  \href{https://doi.org/10.1103/PhysRevLett.108.111601}{\emph{Phys. Rev. Lett.}
  {\bfseries 108} (2012) 111601}
  [\href{https://arxiv.org/abs/1111.5206}{{\ttfamily 1111.5206}}].

\bibitem{Buccioni:2017yxi}
F.~Buccioni, S.~Pozzorini and M.~Zoller, \emph{{On-the-fly reduction of open
  loops}}, \href{https://doi.org/10.1140/epjc/s10052-018-5562-1}{\emph{Eur.
  Phys. J.} {\bfseries C78} (2018) 70}
  [\href{https://arxiv.org/abs/1710.11452}{{\ttfamily 1710.11452}}].

\bibitem{Buccioni:2019sur}
F.~Buccioni, J.-N.~Lang, J.M.~Lindert, P.~Maierh{\"o}fer, S.~Pozzorini,
  H.~Zhang et~al., \emph{{OpenLoops 2}},
  \href{https://doi.org/10.1140/epjc/s10052-019-7306-2}{\emph{Eur. Phys. J.}
  {\bfseries C79} (2019) 866}
  [\href{https://arxiv.org/abs/1907.13071}{{\ttfamily 1907.13071}}].

\bibitem{Jezo:2018yaf}
T.~Je\v{z}o, J.M.~Lindert, N.~Moretti and S.~Pozzorini, \emph{{New NLOPS
  predictions for $\boldsymbol{t \bar{t} +b}$ -jet production at the LHC}},
  \href{https://doi.org/10.1140/epjc/s10052-018-5956-0}{\emph{Eur. Phys. J.}
  {\bfseries C78} (2018) 502}
  [\href{https://arxiv.org/abs/1802.00426}{{\ttfamily 1802.00426}}].

\bibitem{Catani_2012}
S.~Catani, L.~Cieri, D.~de~Florian, G.~Ferrera and M.~Grazzini,
  \emph{Vector-boson production at hadron colliders: hard-collinear
  coefficients at the nnlo},
  \href{https://doi.org/10.1140/epjc/s10052-012-2195-7}{\emph{The European
  Physical Journal C} {\bfseries 72} (2012) }.

\bibitem{Davies:1984hs}
C.T.H.~Davies and W.J.~Stirling, \emph{{Nonleading Corrections to the Drell-Yan
  Cross-Section at Small Transverse Momentum}},
  \href{https://doi.org/10.1016/0550-3213(84)90316-X}{\emph{Nucl. Phys.}
  {\bfseries B244} (1984) 337}.

\bibitem{Dicus_1999}
D.~Dicus, T.~Stelzer, Z.~Sullivan and S.~Willenbrock, \emph{Higgs-boson
  production in association with bottom quarks at next-to-leading order},
  \href{https://doi.org/10.1103/physrevd.59.094016}{\emph{Physical Review D}
  {\bfseries 59} (1999) }.

\bibitem{Bal_zs_1999}
C.~Balázs, H.-J.~He and C.-P.~Yuan, \emph{Qcd corrections to scalar production
  via heavy quark fusion at hadron colliders},
  \href{https://doi.org/10.1103/physrevd.60.114001}{\emph{Physical Review D}
  {\bfseries 60} (1999) }.

\bibitem{Harlander_2003}
R.V.~Harlander and W.B.~Kilgore, \emph{Higgs boson production in bottom quark
  fusion at next-to-next-to-leading order},
  \href{https://doi.org/10.1103/physrevd.68.013001}{\emph{Physical Review D}
  {\bfseries 68} (2003) }.

\bibitem{Hamilton:2012rf}
K.~Hamilton, P.~Nason, C.~Oleari and G.~Zanderighi, \emph{{Merging H/W/Z + 0
  and 1 jet at NLO with no merging scale: a path to parton shower + NNLO
  matching}}, \href{https://doi.org/10.1007/JHEP05(2013)082}{\emph{JHEP}
  {\bfseries 05} (2013) 082} [\href{https://arxiv.org/abs/1212.4504}{{\ttfamily
  1212.4504}}].

\bibitem{deFlorian:2016spz}
{\scshape LHC Higgs Cross Section Working Group} collaboration, \emph{{Handbook
  of LHC Higgs Cross Sections: 4. Deciphering the Nature of the Higgs Sector}},
   \href{https://arxiv.org/abs/1610.07922}{{\ttfamily 1610.07922}}.

\bibitem{NNPDF:2021njg}
{\scshape NNPDF} collaboration, \emph{{The path to proton structure at 1\%
  accuracy}}, \href{https://doi.org/10.1140/epjc/s10052-022-10328-7}{\emph{Eur.
  Phys. J. C} {\bfseries 82} (2022) 428}
  [\href{https://arxiv.org/abs/2109.02653}{{\ttfamily 2109.02653}}].

\bibitem{Buckley:2014ana}
A.~Buckley, J.~Ferrando, S.~Lloyd, K.~Nordstr{\"o}m, B.~Page, M.~R{\"u}fenacht
  et~al., \emph{{LHAPDF6: parton density access in the LHC precision era}},
  \href{https://doi.org/10.1140/epjc/s10052-015-3318-8}{\emph{Eur. Phys. J.}
  {\bfseries C75} (2015) 132}
  [\href{https://arxiv.org/abs/1412.7420}{{\ttfamily 1412.7420}}].

\bibitem{Dulat:2015mca}
S.~Dulat, T.-J.~Hou, J.~Gao, M.~Guzzi, J.~Huston, P.~Nadolsky et~al.,
  \emph{{New parton distribution functions from a global analysis of quantum
  chromodynamics}},
  \href{https://doi.org/10.1103/PhysRevD.93.033006}{\emph{Phys. Rev. D}
  {\bfseries 93} (2016) 033006}
  [\href{https://arxiv.org/abs/1506.07443}{{\ttfamily 1506.07443}}].

\end{thebibliography}\endgroup
\end{document}